\documentclass[journal,twocolumn,12pt,final]{IEEEtran}
\ifCLASSINFOpdf
  \usepackage{amsmath}
  \usepackage[pdftex]{graphicx}
  \graphicspath{{../jpeg/}}
  \DeclareGraphicsExtensions{.jpg}
  \usepackage{listings}
  \usepackage[justification=centering]{caption}
  \usepackage{lipsum}
  \DeclareCaptionFormat{listing}{#1#2#3}
\usepackage{textcomp}
  \lstnewenvironment{code}[1][]%
    {\noindent\minipage{\linewidth}\medskip 
     \lstset{basicstyle=\ttfamily\footnotesize,frame=single,#1}}
    {\endminipage}
\else
\fi
\begin{document}
\title{High-Level System Design of IEEE 802.11b Standard-Compliant Link Layer for MATLAB-based SDR}
\author{Ramanathan~Subramanian,~\IEEEmembership{Member,~IEEE,}
	Benjamin~Drozdenko,~\IEEEmembership{Member,~IEEE,}
	Eric~Doyle,~\IEEEmembership{Member,~IEEE,}
        Rameez~Ahmed,~\IEEEmembership{Member,~IEEE,}
        Miriam~Leeser,~\IEEEmembership{Senior~Member,~IEEE,}
        and ~Kaushik~R.~Chowdhury,~\IEEEmembership{Senior~Member,~IEEE}% <-this % stops a space
\thanks{Department of Electrical and Computer Engineering, Northeastern University, Boston,
MA, 02115 USA e-mail: rsubramanian@coe.neu.edu, bdrozdenko@coe.neu.edu, doyle.e@husky.neu.edu, rarameez@gmail.com, mel@coe.neu.edu, krc@ece.neu.edu}% <-this % stops a space
%\thanks{Manuscript received December 1, 2015.}
}
\maketitle

\begin{abstract}
Software defined radio (SDR) allows unprecedented levels of flexibility by transitioning the radio communication system from a rigid hardware platform to a more user-controlled software paradigm. However, it can still be time consuming to design and implement such SDRs as they typically require thorough knowledge of the operating environment and a careful tuning of the program. In this work, our contribution is the design of a bidirectional transceiver that runs on the commonly used USRP\textsuperscript{\textregistered{}} platform and implemented in MATLAB\textsuperscript{\textregistered{}} using standard tools like MATLAB Coder\textsuperscript{\texttrademark{}} and MEX to speed up the processing steps. We outline strategies on how to create a state-action based design, wherein the same node switches between transmitter and receiver functions. Our design allows optimal selection of the parameters towards meeting the timing requirements set forth by various processing blocks associated with a DBPSK physical layer and CSMA/CA/ACK MAC layer so that all operations remain functionally compliant with the IEEE 802.11b standard for the 1 Mbps specification. The code base of the system is enabled through the Communications System Toolbox\textsuperscript{\texttrademark{}} and incorporates channel sensing and exponential random back-off for contention resolution. The current work provides an experimental testbed that enables creation of new MAC protocols starting from the fundamental IEEE 802.11b standard. Our design approach guarantees consistent performance of the bi-directional link, and the three node experimental results demonstrate the robustness of the system in mitigating packet collisions and enforcing fairness among nodes, making it a feasible framework in higher layer protocol design.
\end{abstract}
% \fi
\IEEEpeerreviewmaketitle
\section{Introduction}

%\begin{table}[!t]
%% increase table row spacing, adjust to taste
%\renewcommand{\arraystretch}{1.3}
% if using array.sty, it might be a good idea to tweak the value of
% \extrarowheight as needed to properly center the text within the cells
%\caption{An Example of a Table}
%\label{table_example}
%\centering
%% Some packages, such as MDW tools, offer better commands for making tables
%% than the plain LaTeX2e tabular which is used here.
%\begin{tabular}{|c||c|}
%\hline
%One & Two\\
%\hline
%Three & Four\\
%\hline
%\end{tabular}
%\end{table}

Software defined radios (SDRs) allow fine-grained control of their operation by executing the processing steps in user-accessible program code~\cite{akyildiz}. This technology forms the building block for applications needing high levels of reconfigurability, such as access points that support multiple wireless standards, or for systems like cognitive radios that incorporate situational intelligence to evolve with the radio frequency (RF) environment~\cite{chowdhury}. For example, in SDRs, the network designer can tune basic elements, such as modulation, spectrum spreading, scrambling, and encoding through software functions, instead of relying on static hardware, thereby allowing unprecedented access to all aspects of the radio operation. However, significant expertise is required to successfully navigate the hardware design, software implementation, wireless standards requirements, and computational timing limitations, which requires specialized training and lengthens time to project completion.  

A basic SDR system is composed of a computer connected to a RF front end capable of receiving and transmitting radio signals. A RF front end requires an antenna suited for specified RF bands of interest, a transceiver chip that is comprised of at least one local oscillator, analog-to-digital converter (ADC), and digital-to-analog converter (DAC), and an interface (e.g. Ethernet cable) that connects the front end to the computer. The computer may have a general purpose processor to process the digital output and programs to realize specialized tasks such as filtering, amplification, and modulation, which have traditionally been implemented in hardware. The design concept of the SDR is advantageous because it reduces the need for special purpose hardware and allows the developer to add new functionality to the radio by modifying the software. The flexibility inherent in the SDR allows for the potential to support many wireless standards, whereas a single hardware transceiver can only support a few or one standard. Hence, the SDR device can be seen as an increasingly affordable alternative.

Any modern wireless standard relies on accurate timing to complete the standards-specified tasks. In SDR, as the received and transmitted signals are represented as arrays of data samples collected by the front-end, software processing contributes to delays. Additionally, when multiple nodes operate in a shared channel, timing issues add to the challenge of ensuring synchronized behavior between multiple nodes. In the absence of hardware clocks, the SDR must devise a means of calculating how much time has elapsed, so that transmission and reception functions are performed at the appropriate intervals. The processing functions and their internal parameters must also be open for change, should a better algorithm be designed, or if no set thresholds may be possible, as is the case in highly challenging environments with variable noise floor. Finally, the software running on the SDR must be structured in a hierarchical manner, so that its functionality can be separated into layers that are compliant with the Open Systems Interconnection (OSI) model. Thus, the base drivers that interface with the RF front-end platform should be abstracted from the physical (PHY) layer functionality, which in turn should be abstracted from the medium access control (MAC) layer logic. In summary, there are many design challenges that must be overcome before a highly customizable SDR platform is made available for general purpose use.

This paper details our approach to realize a SDR platform using commonly available tools. We believe that true and repeatable systems-level research is only possible when a commonly used processing environment is used in conjunction with affordable SDR hardware. This motivates our choices for basing our work on MATLAB software and Ettus USRP\textsuperscript{\textregistered{}} N210 hardware \cite{ettus}. Our approach introduces a novel methodology for an implementation starting at the USRP hardware driver (UHD) and building progressively up the protocol stack. To facilitate quick deployment, it includes an initialization script for the setting and tuning of the reconfigurable parameters at the physical layer based on the specific channel measurements at the chosen experimental site. Importantly, it complies with the processing definitions in the IEEE 802.11b specification, though hardware limitations increase the time to completion of the entire transmission/reception cycle compared to an off-the-shelf hardware-only Network Interface Card. 

%\subsection{Our Contributions}
Our approach advances the state of the art and contributes to the research community in the following ways: \\
%\begin{enumerate}
\emph{\bf Standards compliant link layer:} Our approach is based on the IEEE 802.11b specifications \cite{ieee80211b}, faithfully modeling the DATA and ACK packet structure, and implements both PHY-layer and MAC-layer protocols. Further, our work provides a testbed to experiment with new MAC protocols starting from the fundamental IEEE 802.11b compliant standard.\\
\emph{\bf State-action based design:} We model our system using a finite state machine (FSM) that transitions only on the clock cycles derived from the USRP clock, allowing for slot-time synchronized operations. In this manner, we eliminate the need for external clocks that would be necessary in a hardware-based design, or interrupts that may be preferable using a real-time operating system.\\
\emph{\bf Design methodology using a common operating environment:} We use the Ettus Research Universal Software Radio Peripheral (USRP) hardware, a radio front end commonly used in wireless research. As the basis for our software design, we use MATLAB R2015b and the Communications System Toolbox Support Package for USRP-based radio \cite{mathworksu}. We use the MATLAB tools such as MATLAB Coder and the MEX interface to provide for acceleration and timing consistency in the execution of system blocks.\\
% is another widely used toolset in academia in industry for modeling algorithms and signal processing systems. We test our design using the Ubuntu operating system, which is free to install. 
\emph{\bf Full parameter flexibility:} Using a software-only approach and parameterizing the most important variables allows the user to reconfigure the system as needed to adapt to changes in its environment. \\
\emph{\bf Publicly available:} Our software is released to the public for research purposes under the GNU Public License (GPL), and is available for download directly from GitHub \cite{github} and MATLAB Central \cite{MATLABCentralFileEx}. The modularity of our code makes it relatively easy to manage and will enable extensibility by the community.\\
%\end{enumerate}

The rest of this paper is organized as follows. In Sec.~\ref{sec_sysarch}, we present the system architecture. We discuss related work on SDR using heterogeneous systems and software platforms in Sec.~\ref{sec_relatedWork}. In Sec.~\ref{sec_sysdesign}, we describe the slot-time synchronized operations around which the state machines for the designated transmitter and receiver are modeled, and we identify the common system blocks. We describe the algorithms implemented for RFFE and preamble detection in the PHY Layer, followed by a discussion on parameter selection and same-frequency channel operation in Sec.~\ref{sec_phyLayer}. The MAC layer design and key algorithms required to implement the CSMA/CA protocol, such as energy detection and random backoff, are described in Sec.~\ref{sec_macLayer}. The experimental setup involving the USRP N210 platform and MathWorks products is given in Section~\ref{sec_exptSetup}. In Sec.~\ref{sec_exptResults}, we undertake a comprehensive performance evaluation of the two node and three node system and establish through the experimental results that the system exhibits fairness. Sec.~\ref{sec_conclusion} concludes the paper.

\section{System Architecture Overview}
\label{sec_sysarch}
\begin{figure}[!t]
\centering
\includegraphics[width=0.42\textwidth]{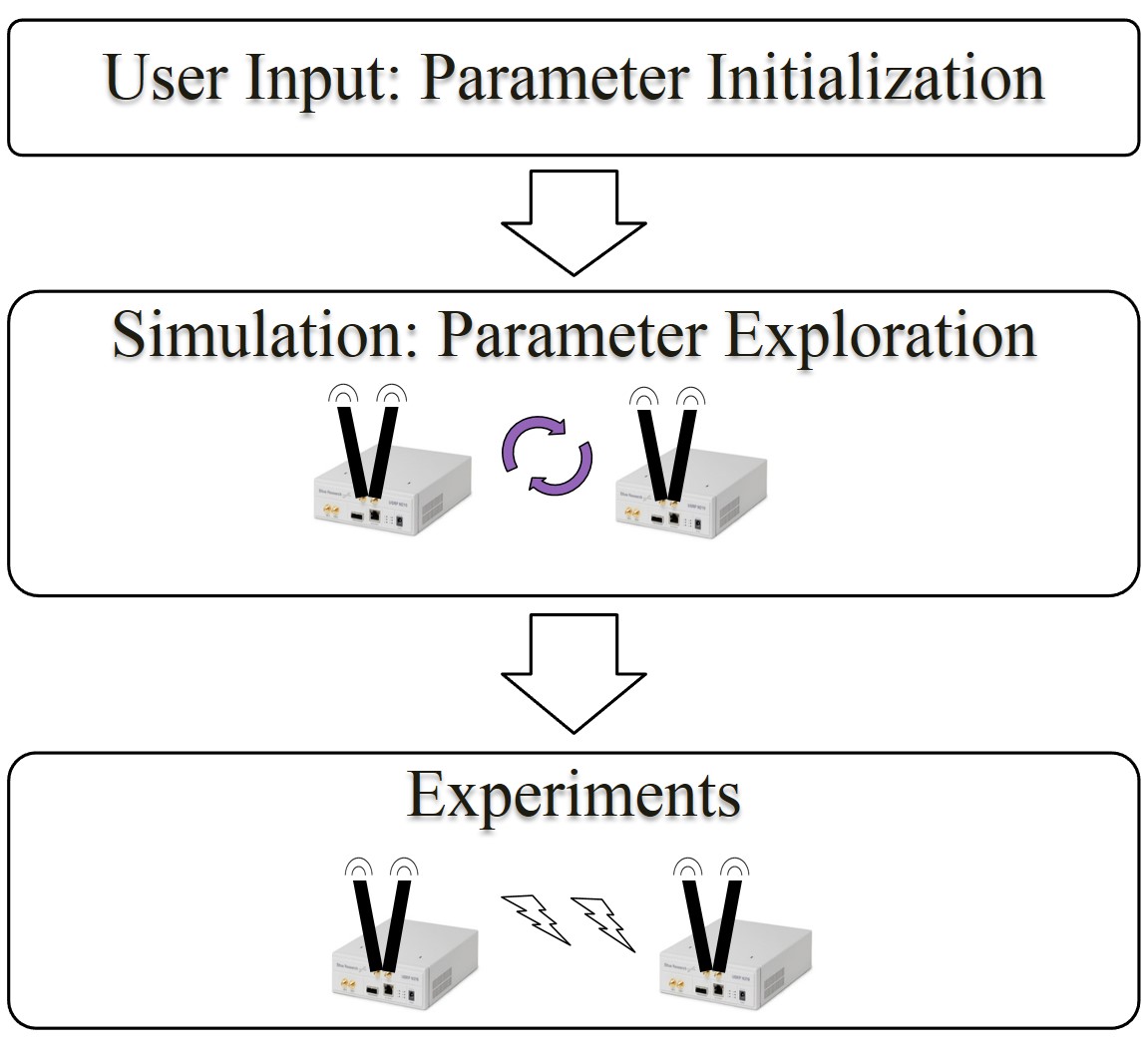}
\caption{System Architecture}
\label{fig_sysarch}
\end{figure}

The operational steps that architect our system are shown in Fig. \ref{fig_sysarch}. In a given SDR pair, we identify clearly the transmitting and receiving node by using the terms designated transmitter (DTx) and designated receiver (DRx). This terminology helps avoid ambiguity in describing a bi-directional transceiver link, where the transmitter must send out its DATA packet and then switch to a receiver role to get the acknowledgement (ACK). Thus, in the discussion ahead, the DTx alternates between its transmit and receive functions, and the DRx alternates between receive and transmit functions.

In the initialization step, the system is preset with recommended parameters and lets the user modify a number of parameters for the entire transceiver chain. The user then, in a simulation-only environment, initiates a parameter exploration stage, where all the \emph{nodes} are virtual and are contained within the same computer. The DTx and DRx codes are executed with the user-supplied parameters as constants, and the code cycles through possible variations in the settings of processing blocks as well as entire algorithms, each time identifying the performance that results from these settings. 

From this data set, the user is presented with a feasible set of parameter settings. These parameter settings result in less than 5\% packet loss at the receiver. This represents the {\em best case} scenario, for it should be noted that further channel outages will be introduced by the actual wireless channel. Once the user selects one of the possible feasible configurations returned by the search, the code is ready for driving the USRPs for over-the-air experiments. 

We adopt the IEEE 802.11b PHY and MAC layer packet structure specifications in our implementation \cite{ieee80211b} \cite{ieee80211}. Our approach collects all the bits in the packet in multiples of 8 octets, which forms one USRP frame. This makes it easy for us to work with the MATLAB system objects (specialized objects required for streaming, henceforth referred to as objects) and with PHY and MAC header fields in the DATA/ACK packet that happen to have sizes that are multiples of 8 octets. Multiple USRP frames will compose the standard-compliant 802.11b packet.

We use differential binary phase shift keying (DBPSK), as the differential component enables us to recover a binary sequence from the phase angles of the received signal at any phase offset, without compensating for phase. In addition, DBPSK requires only coarse frequency offset compensation, without any closed-loop techniques. If residual frequency offset is much less than DBPSK symbol rate, then the bit error rate (BER) approaches theoretical values \cite{drozdenko}.

\section{Related Work}
\label{sec_relatedWork}
\subsection{SDR Software Platforms}
Specialized software is needed to effectively work with the SDR systems and perform the signal processing tasks needed to instantiate wireless communications, such as modulation, preamble detection, encoding, and filtering. GNU Radio is one of the most widely used SDR programs, owing to the fact that it's open source, hardware-independent, and modifiable \cite{gnuradio}. Its GUI, GNU Radio Companion, allows the user to build block diagrams to represent complex encoding and decoding schemes. Modules are built in C++, ordering of components performed in Python, and connections are made using SWIG. Built-in modules allow the user to perform various types of modulation (e.g. GMSK, PSK, QAM, OFDM) and error-correcting codes (e.g. Reed Solomon, Viterbi, turbo). The Software Communications Architecture (SCA) is another open-source, HW-independent framework that models SDR components using data flow diagrams. It is also written using C++ and Python, but intra-block message-passing is accomplished using Common Object Request Broker Architecture (CORBA) middleware. Different software blocks are graphically represented using Unified Modeling Language (UML). The OSSIE software effects an SDR using the SCA framework for interaction with the USRP board \cite{gonzalez}. OSSIE provides a GUI to enable the designer to create new waveforms, add new signal processing and modulation routines, and generate the C++/Python code for SCA-CORBA interactions.

\subsection{SDR on Heterogeneous Systems}
There are existing SDR projects implemented on heterogeneous systems that make use of a combination of hardware components to handle computing tasks, including digital signal processors (DSPs), application-specific integrated circuits (ASICs), and field-programmable gate arrays (FPGAs). \cite{simon} describes an SoC design for placing transceiver components, including RF receivers at 2~GHz and 5~GHz, a voltage controlled oscillator (VCO), and a baseband filter. \cite{kim} proposes a hardware architecture for an embedded software modulation/demodulation (modem) platform, implementing IEEE 802.11a PHY using the Altera Stratix II FPGA and S3C2410 ARM processor. \cite{jiao} realizes BX501 components on an ASIC and hardware modules for MAC-layer control on FPGA in Verilog. 

In addition, there are SDR projects that are implemented in both hardware and software on a platform that comprises both processor and FPGA, and this often includes many custom-made components. WARP is scalable, extensible programmable wireless platform produced by Rice University to prototype advanced wireless networks \cite{warp}. It combines a MAX2829 RF transceiver, high-performance programmable hardware Xilinx Virtex-4 FPGA board, and an open-source repository of reference designs and support materials. This platform has been used to build, among many other things, a full duplex IEEE 802.11 network with OFDM and a MAC protocol \cite{duarte}, and a distributed energy-conserving cooperation MAC protocol for MIMO performance improvements \cite{hunter}. USC SDR presents a wireless platform to remove bottlenecks from current SDR architectures \cite{balan}. It combines Xilinx VC707 PCI FPGA development boards with self-sufficient radio front-end daughterboards to make a MIMO testbed, using the FPGA Mezzanine Card (FMC) connection. Real-time SW architecture allows user programs to perform signal processing tasks, PHY- and MAC-layer algorithms. The Sora soft-radio stack combines a Radio Control Board (RCB) with a multi-core CPU.  The RCB that consists of a Virtex-5 FPGA, PCIe-x8 interface, and 256 MB of DDR2 SDRAM \cite{tan}. Microsoft Research built the SoftWiFi Demo radio system to interoperate with 802.11a/b/g NICs, and it uses a company-proprietary language for SDR description. 

There are other SDR projects that are implemented using Xilinx Zynq SoC, utilizing both the PS/ARM processor and PL/FPGA fabric. Iris uses XML description to link together components to form a full radio system \cite{vandebelt}. Components are run within an engine, which could be either a PS processor core or PL logic fabric. It's tested using OFDM for video transmission. GReasy presents a GNU radio version for Xilinx Zynq, using Tflow to instantly program FPGA fabric \cite{marlow}. \cite{ozgul} uses Zynq SoC to implement digital pre-distortion algorithm (DPD), which mitigates the effects of power amplifier (PA) nonlinearity in wireless transmitters, something required for 3G/4G base stations. This uses Vivado HLS to design the PL component and receives up to 7X speedup from HW acceleration. \cite{dobson} proposes a scalable cluster of Zynq ZC702 boards, controlled by a Zedboard that acts as a task mapper to partition data flows across the Zynq FPGAs and ARM cores. tFlow rapid reconfiguration software was used to build FPGA images from a library of pre-built modules. 

\cite{collins} describes an SDR-based testbed that implements a full-duplex OFDM physical layer and a CSMA link layer using MATLAB R2013a, MATLAB Coder on USRP-N210 and USRP2 hardware. The IEEE 802.11a based PHY layer, incorporates timing recovery, frequency recovery, frequency equalization, and error checking. The CSMA link layer involves energy detection based carrier sensing and stop-and-wait ARQ. It outlines some strategies in establishing bidirectional communications. However, this approach involves additional development efforts to improve speed and enable full-duplex operation.

The above platforms make for capable choices in terms of performance. However, our choice of the operating environment was motivated by the price point, which is why we chose to use the combination of USRP N210 hardware and MATLAB software towards link layer implementation. So far there has been little support for MATLAB in the existing SDRs and, in this regard, our framework allows for quick development of new higher layer protocol design. In addition, our software-only infrastructure allows for full flexibility of parameter choices, an option not available to many other SDR platforms. 

\section{State-action based System Design}
\label{sec_sysdesign}
Our approach involves first designing a number of (i) state diagrams to reflect the logical and time-dependent operational steps of our system, and (ii) block diagrams to reflect the sequential order of operations. Furthermore, we structure the MATLAB code in a way that enables slot-time synchronized operations. For the implementation, we use MATLAB Coder to generate the MEX functions for the USRP objects on an Ubuntu 64-bit platform that serves as the host computer for the USRPs. 

Since the underlying code in a MEX function is written in C, it is generally faster than the interpreted MATLAB. The speed-up in performance can vary depending on the application. In our case, we preferred the MEX interface because it can enforce a consistent processing time per frame. The interpreted MATLAB, unlike the MEX, lacks this ability because it exhibits significant deviation from the desired timing. In addition, time-sensitive operations such as frequency offset compensation, show speed improvement using MEX.

\begin{figure}[!t]
\centering
\includegraphics[width=3in]{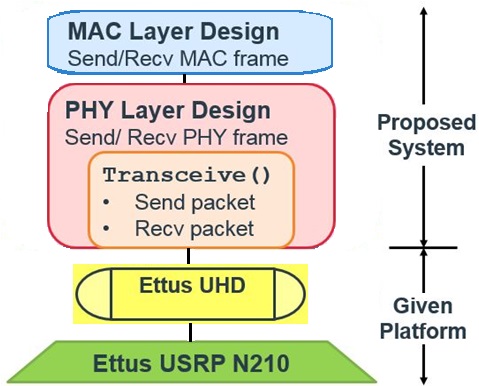}
\caption{System Methodology}
\label{fig_meth}
\end{figure}

Our system design builds upon an already-defined platform, the USRP, produced by a well-known platform supplier, Ettus Research \cite{ettus}. The communication between the USRP and host computer is established in MATLAB using the Communications System Toolbox (CST) USRP Radio support package, which acts as a wrapper for the Ettus USRP Hardware Driver (UHD) drivers.  Identifying the manner in which the RF samples are transported between the USRP and a calling function defines the manner in which we must build the physical (PHY) layer, as illustrated in Fig. \ref{fig_meth}.

The UHD transfer of a frame of samples to a transmit buffer is performed as soon as it is requested while the UHD retrieval of a frame from a receive buffer has to wait until the next rising edge of a clock cycle before trying to retrieve again. The most common undesirable behaviors that can occur are \emph{underflow} and \emph{overflow}. \emph{Underflow} occurs when the radio requests for a frame of data from the transmit buffer, but the host is not yet ready to provide it. \emph{Overflow} occurs when the receive buffer becomes full and buffered data must be overwritten. 

In this regard, we define real-time operation over the course of an entire DATA-ACK packet exchange using equation (\ref{eqn_realTime}) below:
\begin{equation}
\label{eqn_realTime}
t_{receive} \leq t_{radio}
\end{equation}
where $t_{radio}$ is the frame time stipulated by the USRP radio\textquotesingle s analog-to-digital converter (ADC) and $t_{receive}$ is the average time to recover any given frame, which includes the time to retrieve a frame from the receive buffer, process the retrieved frame to decode it into the corresponding bits, and other memory and conditional operations.

Essentially, we operate in real-time if we meet the timing deadline set forth by equation (\ref{eqn_realTime}). Such an operation will guarantee a stable, basic bi-directional link that shows no sign of any undesirable system behavior, such as buffer underflow or buffer overflow. A MAC protocol that effectively schedules packet transmissions reduces the potential for packet collisions and buffer overflow, thereby decreasing packet errors.

\subsection{Slot-time synchronized operations}\label{sec_transceive}
Any IEEE 802.11-based wireless transceiver implementation must have the ability to perform operations based on some slot-based timing. Performing such slot-time synchronized operations will let us realize time-sensitive functions, for example, make a node wait for a backoff (BO) duration before sending a DATA packet.

Interpreted MATLAB or any other software that runs on the host computer may have trouble performing such operations in this manner, even by actively waiting. For this reason, we rely on the USRP for our timing. Using the value for USRP interpolation/decimation defined in Section \ref{sec_paramconst}, we can calculate the slot time. Then, we write our while loop in the main program so that it calls the transceive function once per loop, running helper functions to prepare data to transmit or process received data based on the active state, as shown in the program code in Listing \ref{lst_mainprogram}.

\begin{code}[caption={Main program calls transceive function},label={lst_mainprogram},language=MATLAB,captionpos=b]
while ~endOfTransmission
  if (state==Tx)
    data2Tx = processData2Tx();
  end
  dataRxd = transceive(data2Tx);
  if (state==Rx)
    processRxdData(dataRxd);
  end
end
\end{code}

At the heart of the transceiver model is the \emph{transceive} function, as shown in Listing \ref{lst_transceive}. By design, \emph{transceive} is called at a constant time interval that we define as a slot time. At each slot time, \emph{transceive} sends and receives a fixed number of samples, which we refer to as a \emph{USRP frame}.

We define a slot time as the smallest unit of time in which our SDR can make a decision. In our design, the frame time is the minimum time our system takes to make a decision and hence, we equate it to the slot time. In this regard, our transceive function performs two actions: it gets a frame from, and puts a frame into the USRP buffers at fixed time intervals \cite{drozdenko}. A data frame is sent or received every slot time and further, the functions we define for processing the received data frame or preparing a new data frame to transmit are intended to complete in less than a slot time to ensure timing accuracy. In practice, we recognize that the processing time for certain frames may exceed the radio time, \emph{$t_{radio}$}, but the recovery time, \emph{$t_{receive}$}, converges to the radio time.

\begin{figure}[!t]
\centering
\includegraphics[width=3.5in]{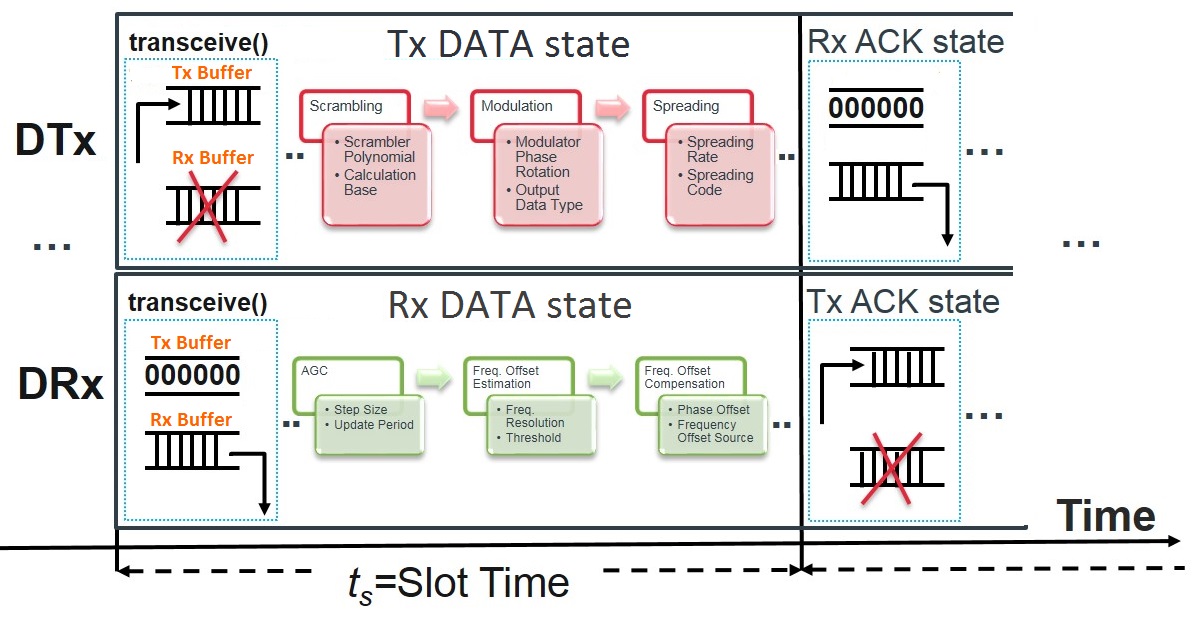}
\caption{Transceive Function Behavior as Defined by Operational State}
\label{fig_trx}
\end{figure}

When a node (either DTx or DRx) enters a transmit state (refer to Fig. \ref{fig_trx}), it transmits the samples in the transmit buffer and ignores all samples in the receive buffer. On the other hand, when a node enters a receive state, it retrieves samples from the receive buffer for processing and puts zeroes in the transmit buffer. This way, we make sure that the samples in the transmit and receive buffer are current and relevant.

\begin{code}[caption={Transceive function code},label={lst_transceive},language=MATLAB,captionpos=b]
function dr = transceive(ft, d2s)
persistent hrx htx;
% Initialize received data variables
dr = complex(zeros(nspf,1));
ns = 0;
% Initialize system objects once
if isempty(hrx)
  hrx = ...; htx = ...;
end
% Flag to release system objects
if ft
  release(hrx); release(htx);
else
  step(htx,d2s);
  while (ns == 0)
    [dr,ns] = step(hrx);
  end
end
\end{code}

The step method of the transmitter object operates in a blocking way as it returns only after the radio accepts the frame to be transmitted. On the other hand, the step method of the receiver object returns right away, hence it is non-blocking. 

The step call of receiver object will return 0 as length of the received frame if there is not enough data in the radio. Once the radio collects enough data, the next step call returns a non-zero length value and the valid data. Since we know the sample rate of the data and the number of samples in a frame, we can calculate how long it takes to get one frame of data from the radio. The while loop blocks the transceive function until a frame of data is received. Therefore, we can use the call duration of this function as our clock source. 
\subsection{Designated Transmitter State Machine}
\begin{figure}[!t]
\centering
\includegraphics[width=3.5in]{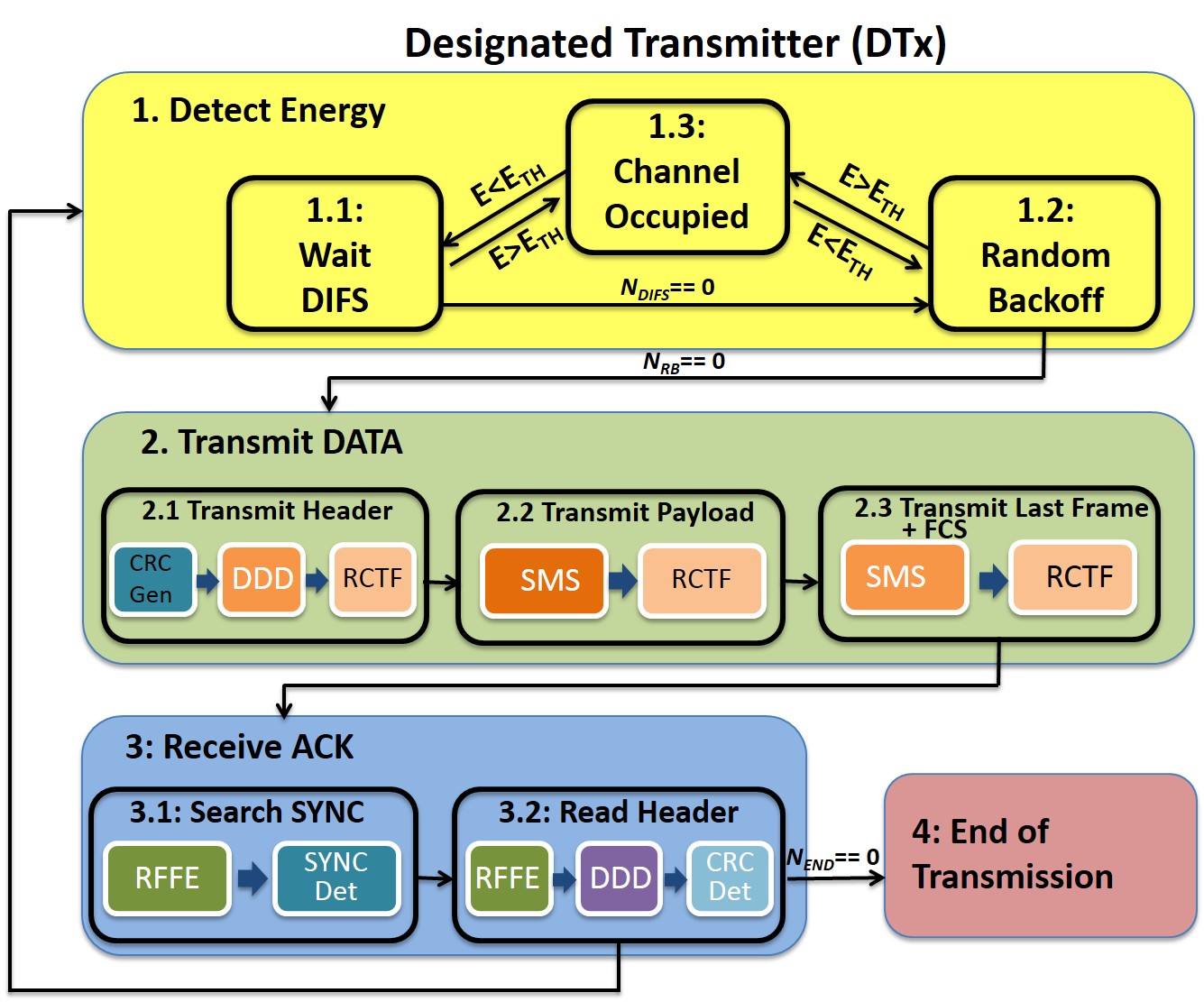}
\caption{States for the Designated Transmitter (DTx)}
\label{fig_dtx}
\end{figure}

In implementing the carrier sense multiple access with collision avoidance (CSMA/CA)-based protocol in the link layer, we identify 4 main states for the DTx, as shown in Fig. \ref{fig_dtx}. Table \ref{table_blocks} identifies the blocks in each substate and is described in detail in Section \ref{sec_sysblocks}.

\begin{table}[!ht]
\renewcommand{\arraystretch}{1.3}
\caption{Substate Operation Combinations}
\label{table_blocks}
\centering
\begin{tabular}{c||c}
\hline
\bfseries Block & \bfseries Block Components\\
\hline SMSRC & Scrambling, Modulation, Spreading, and\\
 & Raised Cosine Transmit Filter (RCTF)\\
\hline RFFE & Radio Frequency Front End: includes\\
 & Automatic Gain Control (AGC), \\
 & Coarse Frequency Offset Estimation (CFOE),\\
 & Frequency Offset Compensation (FOC),\\
 & and Raised Cosine Receive Filter (RCRF)\\
\hline PD & Preamble/SYNC Detection: \\
 & Find SYNC in Rx'd USRP frames \\
\hline DDD & Despreading, Demodulation, and Descrambling\\
\hline
\end{tabular}
\end{table}
\subsubsection{Detect Energy}
At the start, a new USRP frame arrives, and gets stored in a receive buffer. The DTx begins to continually sense energy in the channel and decides to transition either into a backoff state or to a transmit state depending on whether or not the channel is busy. It first waits for a DCF interframe spacing (DIFS) duration and then waits for a random amount of time that is chosen uniformly from a progressively increasing time interval. Only when the channel is free does the DTx decrement the chosen random backoff time; otherwise, it stalls. Only when the backoff time counts down to zero does the DTx attempt to transmit. 
\subsubsection{Transmit DATA}
Upon entering this state, the DTx prepares the DATA packet and then, by calling the transceive function continually, places it in the transmit buffer of the USRP which then gets transmitted over the air. After transmitting the DATA packet, two possibilities exist. The transmission is successful with the reception of an ACK, or the transmission is not successful due to packet collision with another DTx.
\subsubsection{Receive ACK}
As soon as the DATA packet is transmitted, the DTx moves into the Receive ACK state, searching and decoding the Physical Layer Convergence Procedure (PLCP) header in the received ACK. If that is successful, the frame control and the address fields are read-out from the subsequent MAC header and checked for accuracy. The DTx then progresses to transmit a new frame and repeats the above mentioned sequence of steps until the last frame is successfully transmitted. On the other hand, if no ACK is received, the packet is considered lost and the DTx backs-off for an increased random backoff time and re-attempts transmission.
\subsubsection{End Of Transmission} 
When there are no more DATA packets left to be transmitted, the DTx reaches the end of transmission (EOT) state.
\subsection{Designated Receiver State Machine}
\begin{figure}[!t]
\centering
\includegraphics[width=3.5in]{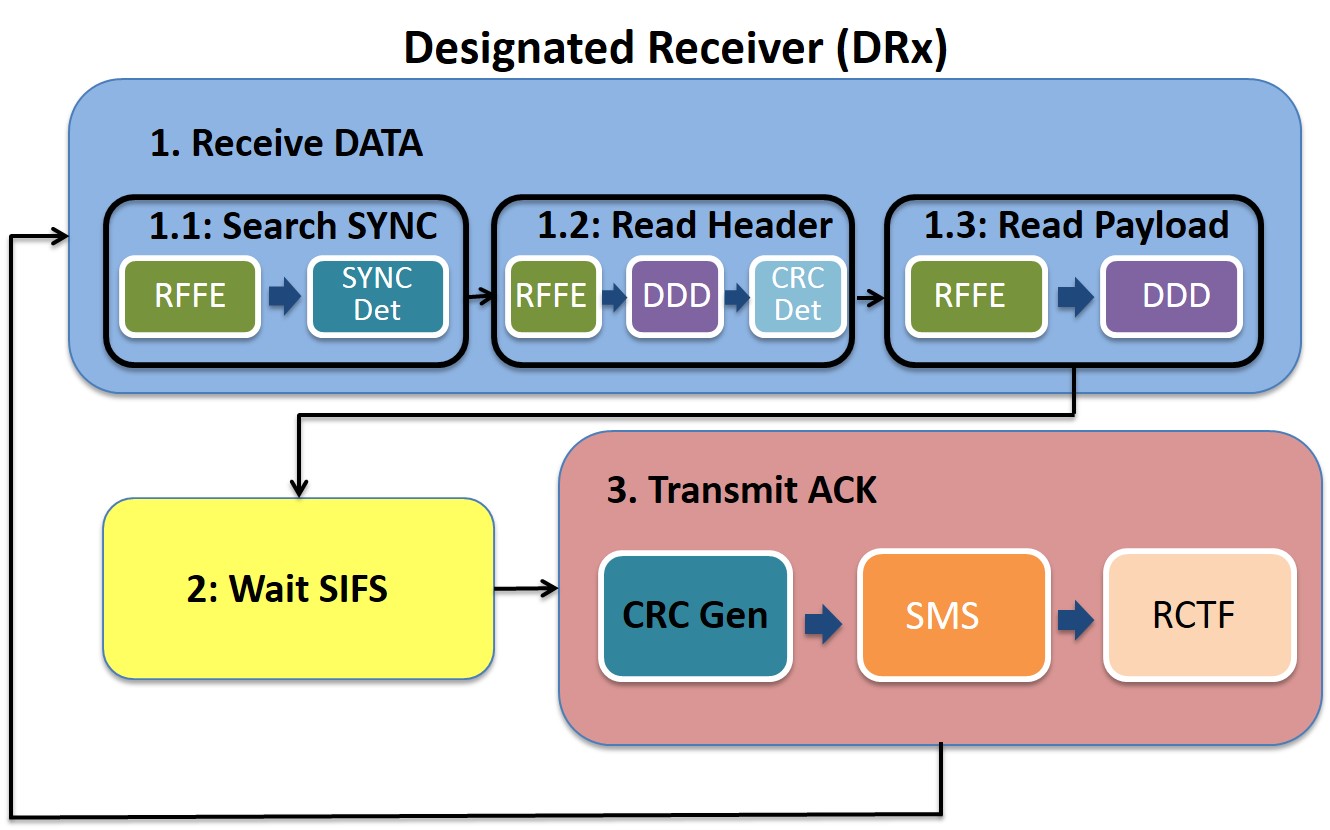}
\caption{States for the Designated Receiver (DRx)}
\label{fig_drx}
\end{figure}

Similarly, we identify 3 main states for the DRx as shown in Fig. \ref{fig_drx}. Unlike the DTx, the DRx does not perform energy detection.
\subsubsection{Receive DATA}
When the DRx successfully detects the Preamble and the Start Frame Delimiter (SFD), it decodes the PHY and MAC header and then progresses to extract the payload. When extracting the last set of payload bits, Frame Check Sequence (FCS) is obtained and checked.
\subsubsection{Wait SIFS}
The DRx waits for a fixed interval of time, referred to as Short Inter-frame Space (SIFS), before sending an ACK packet post reception of the DATA packet. 
\subsubsection{Transmit ACK}
The DRx sends out an ACK addressed to the DTx when it successfully retrieves all the payload bits.
\subsection{System Blocks}\label{sec_sysblocks}
Within each of the substates in the FSM diagrams (Figs. \ref{fig_dtx} and \ref{fig_drx}), there are sequential operations that need to be performed.  In order to simplify the logic of which operations must be performed in each state, we define a number of \emph{blocks} to comprise the most common operations, as shown in Table \ref{table_blocks}. Identifying the grouping of blocks with the related substates helps better organize and restructure the implemented code. 

In each substate of DTx state 2 (Tx) and DRx state 2 (Tx ACK), SMSRC is performed prior to each \emph{transceive} (send and receive operation). In DTx substate 3.1 and DRx substate 1.1, RFFE and PD are performed after each transceive. In DTx substate 3.2 and DRx substates 1.2, RFFE and DDD are performed after each transceive. 
\section{PHY Layer Algorithms}
\label{sec_phyLayer}
\subsection{RF Front End Algorithms}
The components in the RFFE block recover a signal prior to preamble detection. These include the automatic gain control (AGC), frequency offset estimation and compensation, and raised cosine filtering. The ordering of these components is an important consideration, and through exhaustive simulations, we found the preceding order to be ideal. The AGC algorithm counters attenuation by raising the envelope of the received signal to the desired level. We chose to use the MATLAB \texttt{comm.AGC} object \cite{mathworksa}.
To accurately estimate the frequency offset between the receiver and the transmitter, we chose to use the \texttt{comm.PSKCoarseFrequencyEstimator} object, which uses an FFT-based-based method, based on equation (\ref{eqn_foe}), and finds the frequency that maximizes the FFT of the squared signal: 
\begin{equation}
\label{eqn_foe}
\textit{\!f}_{\!o\!f\!f\!set} = \arg\max_{f} \mathcal{F}\{x^2\}
\end{equation}
where $x$ is the signal, $\mathcal{F}$ denotes the Fast Fourier Transform (FFT), and \ $\textit{\!f}_{\!o\!f\!f\!set}$ is the frequency offset.  
\subsubsection{Speeding up the RFFE block}
From our initial experiments, we know that a frequency resolution (on the order of 1-10~Hz) is necessary in order to do preamble detection accurately. Setting such a low frequency resolution takes too long to execute with a sample rate of 200~kHz, or 200,000 samples per sec. For this reason, we decided to decimate the signal by a factor of 22 (the RCRF factor times the spreading rate) before CFOE, which is, in essence, an FFT. After decimation, we experimented with raising the CFOE's frequency resolution by an order of magnitude to 10-100~Hz, and determined that it is accurate up to 100~Hz and meets the timing guidelines set by radio time.

We employ a FIR Decimator step, as shown in Listing~\ref{lst_fir}, that enables us achieve an order of magnitude reduction in RFFE block execution time. In essence, we are able to get enough frequency estimation accuracy with reduced sample rate (hence the use of decimation) and 100~Hz frequency resolution, which requires much less processing power than full frame higher resolution estimates. 
\\
\\
\begin{code}[caption={RFFE Decimation Method},label={lst_fir},language=MATLAB,captionpos=b]
(1) dsp.FIRDecimator('DecimationFactor',22);
(2) comm.PSKCoarseFrequencyEstimator(
	'Algorithm','FFT-based', ...
        'FrequencyResolution',cef,...
	'ModulationOrder',2,...
	'SampleRate',(2e5/22));
\end{code}
\subsection{Preamble Detection Algorithms}
The IEEE 802.11b preamble is a sequence of all one bits that undergoes scrambling. Since the scrambling phase is not known, and the received signal is correlated to the zero phase scrambled sequence, the maximum correlation position may not be the synchronization position. Therefore, the standard provides Start Frame Delimiter (SFD), to fine tune the synchronization time.\par Preamble detection (PD) is performed in two stages. In the first stage, we perform a cross-correlation of the received complex data after raised cosine filtering with the expected real preamble to get an estimate of where the preamble starts, giving the so called synchronization delay. Finally, in the second stage, we look for the SFD immediately after the preamble in the descrambled bit stream. If it is not in the expected place, we perform a cross-correlation on a window of descrambled frame samples to the left and right to further fine-tune the synchronization delay.
\subsubsection{Optimization of Preamble Detection}
\begin{figure}[!t]
\centering
\includegraphics[width=3.5in]{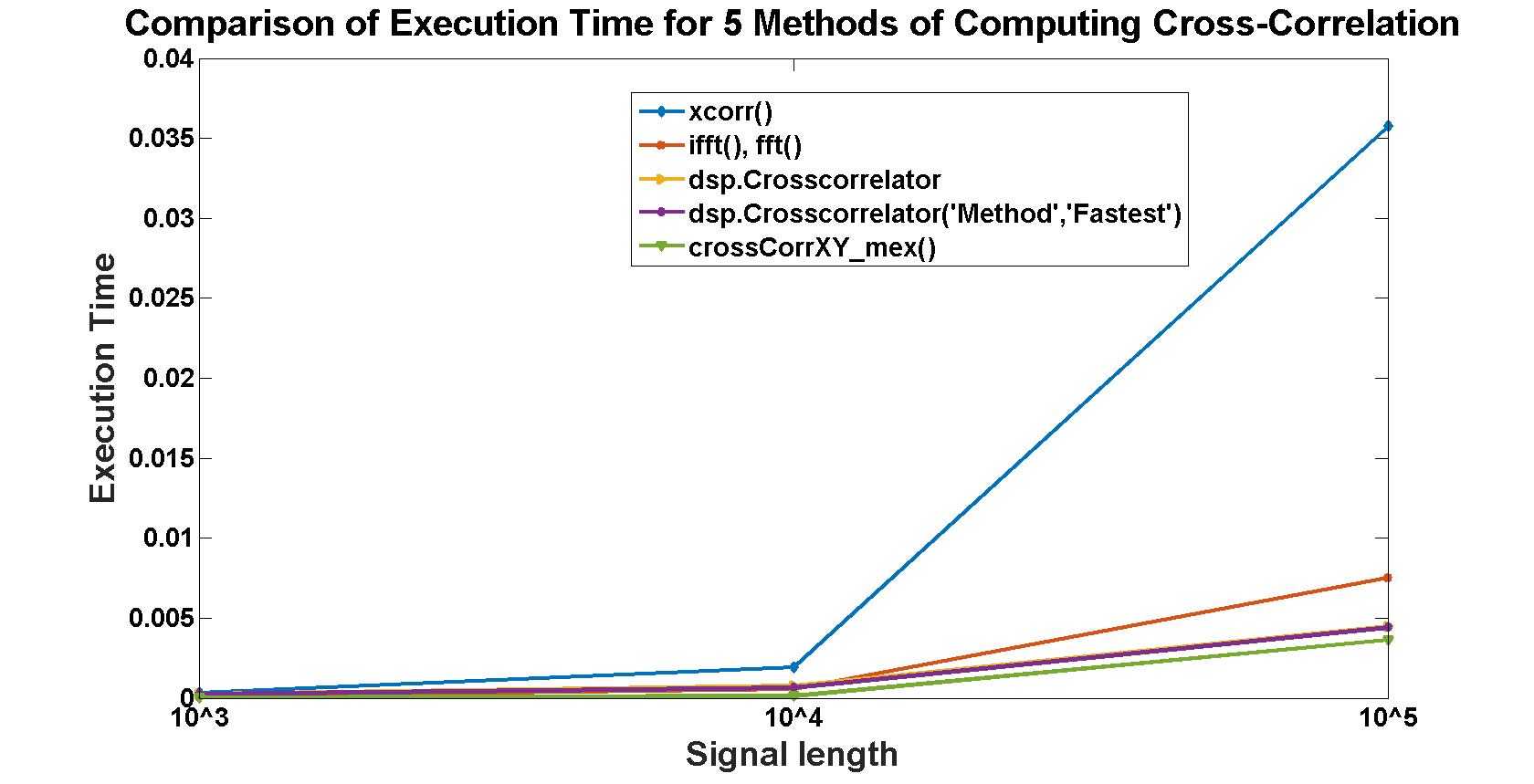}
\caption{Comparison of Execution Time for 5 Methods of Computing Cross-Correlation}
\label{fig_5MethodsCrossCorr}
\end{figure}

Detecting the Preamble fast and with high accuracy is critical to the speed at which the nodes can reliably exchange DATA/ACK packets.
In one implementation, we exploit the property of the cross-correlation of two real signals in the frequency domain to compute the same (i.e. the point-wise product of the Fourier transform of the two signals), followed by an inverse Fourier transform resulting in the cross-correlation of the two signals. Since one of the signals is the expected preamble, its Fourier transform can be pre-computed and loaded into the workspace during run-time.\\
We experimented with several MathWorks utilities to compute cross-correlation faster (e.g. \texttt{dsp.CrossCorrelation} object, \texttt{xcorr} function).\\
We determined the version of \texttt{dsp.Crosscorrelator('method', 'fastest')} compiled using MEX to be the fastest among all the candidate methods for computing cross-correlation with increasing signal lengths, as shown in Fig. \ref{fig_5MethodsCrossCorr}. It is important to note that although we operate with signal lengths on the order of $10^3$, preamble detection is a frequent operation, so savings in time add up quickly.\\
We declare packet detection only if the second stage finds a perfect match for the SFD. This approach greatly minimizes false packet detections.
\subsection{Parameter Selection}
The initialization step described in Section \ref{sec_sysarch} lets us carefully choose a number of design parameters (see table \ref{table_params}).
\begin{table}[!t]
\renewcommand{\arraystretch}{1}
\caption{Important Parameters}
\label{table_params}
\centering
\begin{tabular}{c|c|c|c|c}
\hline
\bfseries Param & \bfseries Block & \bfseries Description & \bfseries Range & \bfseries Tunable \\
\hline $R_{i}, R_{d}$ & USRP & USRP Interpolation & 500 & No\\
  &  & Decimation Factor & & \\
\hline $L_{f}$ & USRP & USRP Frame & 64 bits & No\\
  &  & Length &  & \\
\hline $L_{p}$ & Frame & \#Octets per 802.11b & 0-2312 & Yes\\
  &  & Packet Payload &  & \\
\hline $K$ & RFFE & AGC Max & 30-60 & Yes\\
  &  & Power Gain &  & \\
\hline $N$ & RFFE & AGC Adaptation & 0.01-0.5 & Yes\\
  &  & Step Size &  & \\
\hline $\Delta f$ & RFFE & Frequency & 1-100~Hz & Yes\\
  &  & Resolution &  & \\
\hline
\end{tabular}
\end{table}
\subsubsection{Constant Parameters for USRP \& IEEE 802.11b Frame}
\label{sec_paramconst}
We recognize parameters that cannot change during packet transmission/reception and have to be fixed. The number of octets in the payload per IEEE 802.11b packet should be maximized to decrease the header overhead. In that case, a large frame size is preferred as it reduces the percentage of overhead processing. On the other hand, the frame size should be minimized to make quick decisions with a small number of samples or bits, unlike a large frame size which increases the frame time, thereby reducing the resolution of time ticks for the system. We chose frame length of 1408 as a well balanced compromise between these two requirements. For this reason, the frame length is left fixed.

The USRP N210 analog-to-digital converter (ADC) operates at a fixed rate of 100~MHz. The USRP interpolation-decimation rates control the rate of transmitting and receiving frames. For example, setting interpolation rate, \emph{$R_{i}$}, and decimation rate, \emph{$R_{d}$}, to 500 ensures that the ADC and DAC convert a sample every 5 $\mu$s, as shown in equation (\ref{eqn_ts}). 
\begin{equation}
\label{eqn_ts}
\begin{split}
t_{sample} = R_i / (100 Msamples/sec) \\\ \ \ \ = 500 / 10^8\ \ \ \ \ \ \ \ \ \ \ \ \ \ \ \ \ \ \ \\= 5\times10^{-6} sec/sample\ \ \ \ \ 
\end{split}
\end{equation}
Setting frame length, \emph{$L_{f}$}, to 1408 samples means that a frame is retrieved by the transceive function every 7.04~ms, as shown in equation (\ref{eqn_tf}). 
\begin{equation}
\label{eqn_tf}
\begin{split}
t_{radio} = L_f \times (R_i / 100 Msamples/sec) \\= 1408 \times (500 / 10^8)\ \ \ \ \ \ \ \ \ \ \ \ \ \ \ \\= 7.04\times10^{-3} sec/frame\ \ \ \ \ \ \ \ 
\end{split}
\end{equation}
Even though our system may take more than 7.04~ms to process a frame every once in a while, the buffers in the USRP receiver prevents the system from overrunning (or lose samples) and the system, on average, stays real-time.
\subsubsection{Tunable Parameters for RFFE Block}
Tunable parameters can change during transception. For example, the AGC adaptation step size controls the convergence speed of a received signal's envelope to the desired level. In other words, it governs the speed of convergence. The frequency offset estimation component's frequency resolution setting is an important design consideration as it is inversely proportional to the FFT length. A lower frequency resolution gives more accurate offset estimates, but with increased computational time. 
\subsection{Same-Frequency Channel Operation}
In a multi-node setting, it is advantageous to operate the transmit and receive links, at the DTx and DRx, in the same band of frequencies. Thus, we set both DTx and DRx to operate at the same center frequency. Unlike different-frequency channel operation, this eliminates the need for repeated switching of transmit and receive center frequencies when transitioning among the energy detection, transmit, and receive states. In addition, it makes for an easier implementation of medium access and contention resolution.

From our initial experiments, we learned that the receive-only port, \emph{RF2}, of the USRP leaks about 7 dBm into the transmit \& receive port, \emph{RF1}. The effect of this leakage causes the DTx to detect the preamble in its own DATA packet while it is waiting for an ACK. We added logic to ensure that the DTx rejects its own DATA packet as soon as it reads the MAC header and does not find the expected ACK frame control sequence.
\section{MAC Layer Design}
\label{sec_macLayer}
We first implement the CSMA/CA protocol that allows the nodes to sense the channel and attempt to transmit packets only when the channel is idle to avoid packet collisions. Then, we modify this base implementation with the standards-specific functions, as described below.\par
\subsection{MAC Overview}
\begin{figure}[!t]
\centering
\includegraphics[width=3.5in]{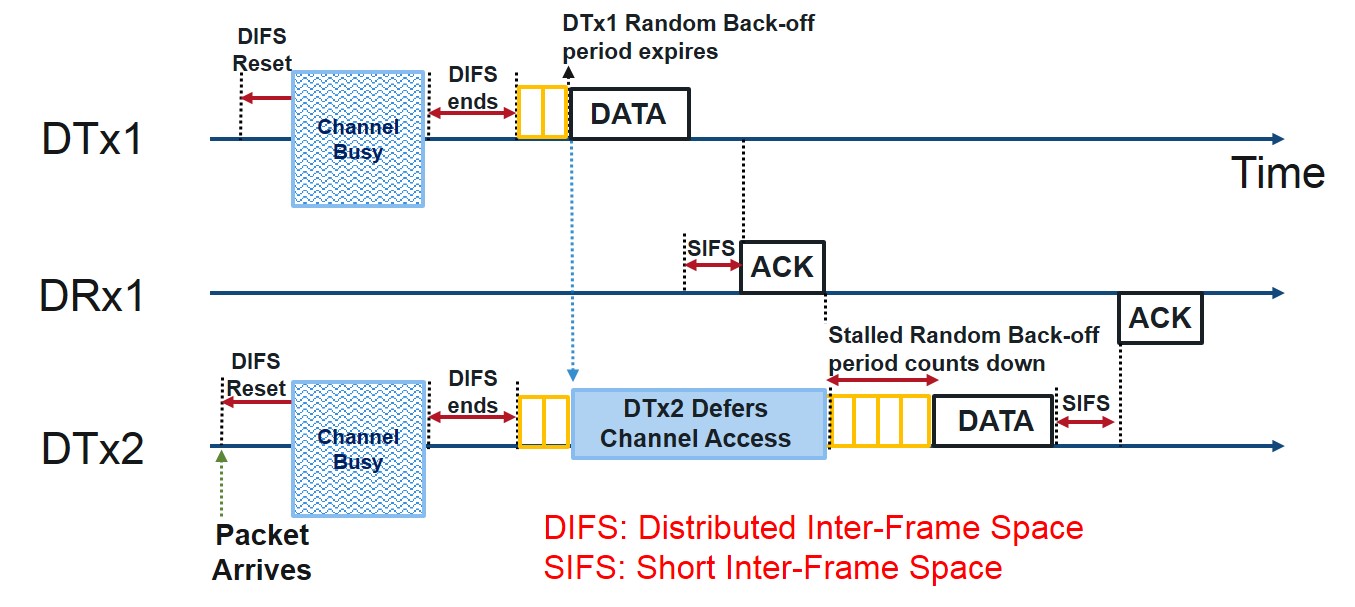}
\caption{CSMA/CA/ACK Timeline Chart - Energy Detection}
\label{fig_CSMA_Chart1}
\end{figure}

\begin{figure}[!t]
\centering
\includegraphics[width=3.5in]{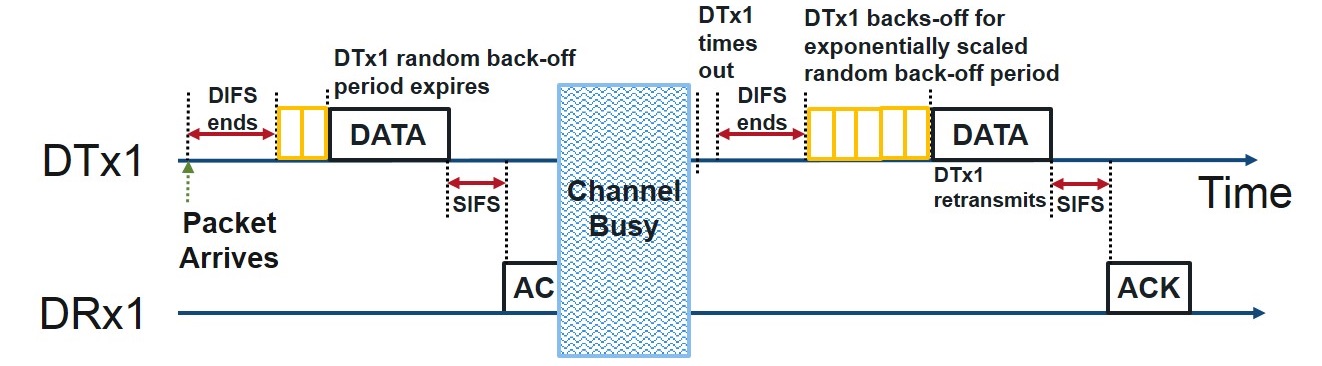}
\caption{CSMA/CA/ACK Timeline Chart - Exponential Random backoff and Retransmission}
\label{fig_CSMA_Chart2}
\end{figure}

Our MAC layer employs the Distributed Coordination Function (DCF) strategy incorporating the CSMA/CA mechanism as it is described in the IEEE 802.11 specification \cite{ieee80211}. Our implementation incorporates the key features of CSMA/CA, namely, 1) carrier sensing via energy detection, 2) DCF interframe spacing (DIFS) duration, and 3) exponential random backoff. An illustration of the overall steps of the operation is shown in Fig. \ref{fig_CSMA_Chart1} and Fig. \ref{fig_CSMA_Chart2}. 

\subsubsection{Energy Detection}
Channel occupancy can be identified by detecting RF energy in the channel. Energy in the channel is computed using equation (\ref{Energy}). 
\begin{equation} \label{Energy}
Energy = \sum_{n=1}^{n=N}|x(n)|^2 \\
\end{equation}
In our implementation, \emph{x}(\emph{n}) represents the samples in the USRP frame retrieved from the receive buffer of the USRP.
\subsubsection{DIFS Period}
The standard specifies that when a packet is prepared by the DTx and ready to be sent to the intended DRx, the DTx must actively listen to the channel for a fixed specified amount of time known as the DIFS period. If during this period, the DTx senses RF signal energy from other transmitting devices (i.e. when the channel is found busy), it defers the transmission and enters a \emph{Channel Occupied} state. In this state, the DTx stays idle as long as the ambient RF energy is above a specified threshold. When the energy drops below the threshold (i.e. the medium is sensed to be free), the DTx resets the DIFS duration and starts counting down again.
\subsubsection{Binary Exponential Random Backoff}
This method of random backoff is used to schedule retransmissions after collisions. Essentially the retransmissions are delayed by an amount of time determined by a minimum contention window, $c_{min}$, and the number of attempts to retransmit the DATA packet. With this increased number of retransmit attempts, the delay can increase exponentially.\par
When the DIFS duration runs out, the DTx transitions to the exponential random backoff state wherein it generates a random backoff delay uniformly chosen in the range [0, \emph{W}-1] where \emph{W} is called the contention window (CW).\par
In correspondence with the IEEE 802.11 standard, time is slotted using a basic time unit which is the time needed to detect the transmission of a packet from any other station. In our implementation, $t_{radio}$ represents the basic time unit for the system, within which we can detect another DTx transmitting.\par
As an example, after $k$ collisions, a random number of slot-times is chosen at random from $[0,2^{k}\mbox{-}1]$ as described in equation (\ref{Random backoff Delay}).
\begin{equation} \label{Random backoff Delay}
\small Random\ Back\mbox{-}o\!f\!f\ Delay = randi[0,2^{k}\mbox{-}1] \times t_{radio}
\end{equation}
The MATLAB \texttt{randi} function picks an integer uniformly at random from the specified interval. In our implementation, we have the option to truncate the exponentiation with a fixed number of retransmits so as to have a ceiling for the Random backoff Delay.
\section{Experimental Setup}
\label{sec_exptSetup}
\begin{figure}[!t]
\centering
\includegraphics[width=3.5in]{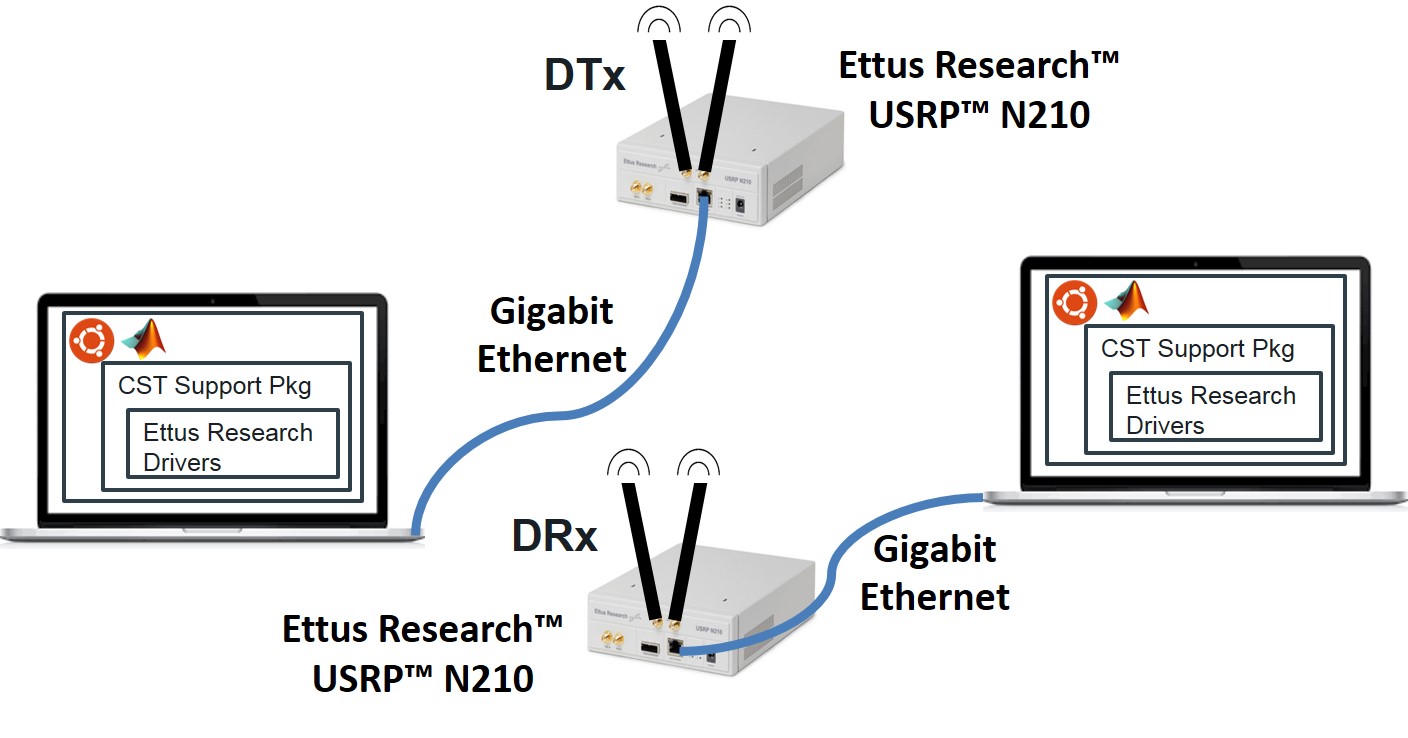}
\caption{Transceiver Hardware Setup}
\label{fig_hwsetup}
\end{figure}

We use the USRP N210 platform \cite{ettus}, as it allows us to define the parameters listed in Section \ref{sec_paramconst}, connect to a PC host using a gigabit Ethernet cable, and to program it using MATLAB \cite{mathworksu}. We use the Ubuntu OS, with send and receive buffer sizes for queues set to ensure that there is enough kernel memory set aside for the network Rx/Tx buffers. We also set the maximum real-time priority for the \texttt{usrp} group to give high thread scheduling priority. This change is made by adding a line to the file \texttt{\textbackslash etc\textbackslash security\textbackslash limits.conf} that sets the \texttt{rtprio} property for the \texttt{@usrp} group to 50. The overall setup is shown in Fig. \ref{fig_hwsetup}.
\subsection{Communications System Toolbox USRP Support Package}
We use the Communications System Toolbox objects for our design \cite{mathworksc}. We used the \texttt{comm.AGC} object and the PSK coarse frequency offset estimator that allows us to work with FFT-based options. These objects facilitate easy generation of C code using MATLAB Coder. Here, the \texttt{comm.SDRuTransmitter} object puts a frame on the USRP transmit buffer, and \texttt{comm.SDRuReceiver} gets a frame from the USRP receive buffer.  However, this approach has some disadvantages, such as a requirement for fixed frame length and single-threaded \texttt{step} methods. 
\subsection{MATLAB Coder}
A number of steps must be taken to make the MATLAB code ready for C code generation using MATLAB Coder. All variables that do not change over the course of the program execution are given a static size and type (including real or complex). All objects are declared as persistent variables as they cannot be passed into MEX functions. The first call to each function tests whether the persistent variable is empty, and initializes each object if true. The transceive and RFFE function code are designed in this manner.
\section{Experiments and Results}
\label{sec_exptResults}
We choose to evaluate our system using a number of experiments. First, we time the reception of DATA packets at the DRx. Next, we time the RFFE block using both interpreted MATLAB and MEX. We then perform a two node experiment, measuring bi-directional link latency and packet error rate. We then profile execution time in the transmitting states. Finally, we perform a three node experiment, measuring previous metrics and goodput. 

In the three node experiment, we address the fairness in our system. Considering two bi-directional links emerging from two DTxs but incident on a DRx helped us to design (within hardware constraints) and demonstrate a stable bi-directional link and allowed us to test the fairness enabled by the MAC protocol in the most simplified way, thereby eliminating the need for further multi-node scenarios. 
Performing more scenarios would require setting up and performing experiments involving multiple nodes and host machines, and would take a large amount of effort. Such an effort would not have helped us in attaining our goal of fairness assessment. 
In addition, we can presume that an increase in the number of DTx nodes would exhibit less fairness because it increases the likelihood of collisions. In this situation, nodes that would collide would also choose to wait for increased backoff periods, which would give other nodes an increased opportunity for transmissions. Additional tests would not be necessary to confirm this hypothesis. 
\subsection{Timing DATA Packet Reception at DRx}
\begin{figure}[!t]
\centering
\includegraphics[width=3.5in]{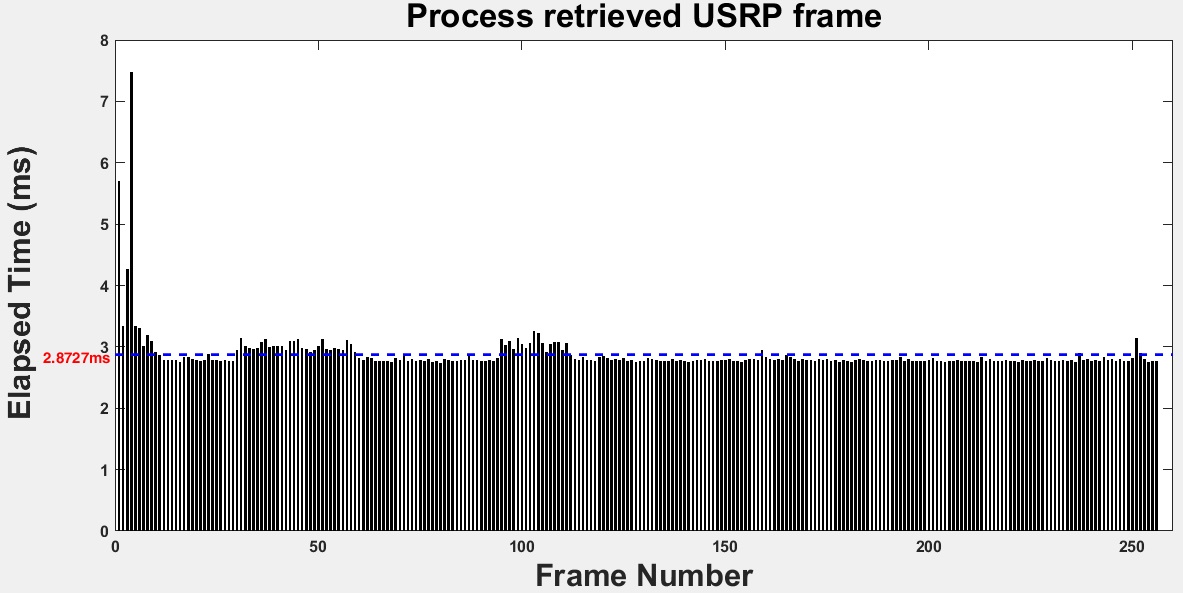}
\caption{Process Time per USRP frame at DRx}
\label{fig_DRxUSRPFrameTime}
\end{figure}

At the DRx, after preamble detection, the elapsed time to process each retrieved USRP frame corresponding to an entire DATA packet is shown in Fig. \ref{fig_DRxUSRPFrameTime}. The dotted line represents the average of all the frame processing times towards a DATA packet reception. 
The DTx sends out a DATA packet that is made up of 258 USRP frames. After recovering the header bits, the DRx retrieves the payload, which is 250.5 USRP frames (2004 octets). Since the Preamble is 128 bits long, it corresponds to 2 USRP frames. Hence, we account for the reception of (258 - 2) = 256 USRP frames in the DATA packet.

The time to process any given frame usually falls below the desired frame time, $t_{radio}$, and is fairly constant at 2.87~ms. The first set of frames have a higher processing time because they consist of the MAC header information that must be resolved (e.g. frame control, MAC address). 
\subsection{RFFE Block Timing}
\begin{figure}[!t]
\centering
\includegraphics[width=3.5in]{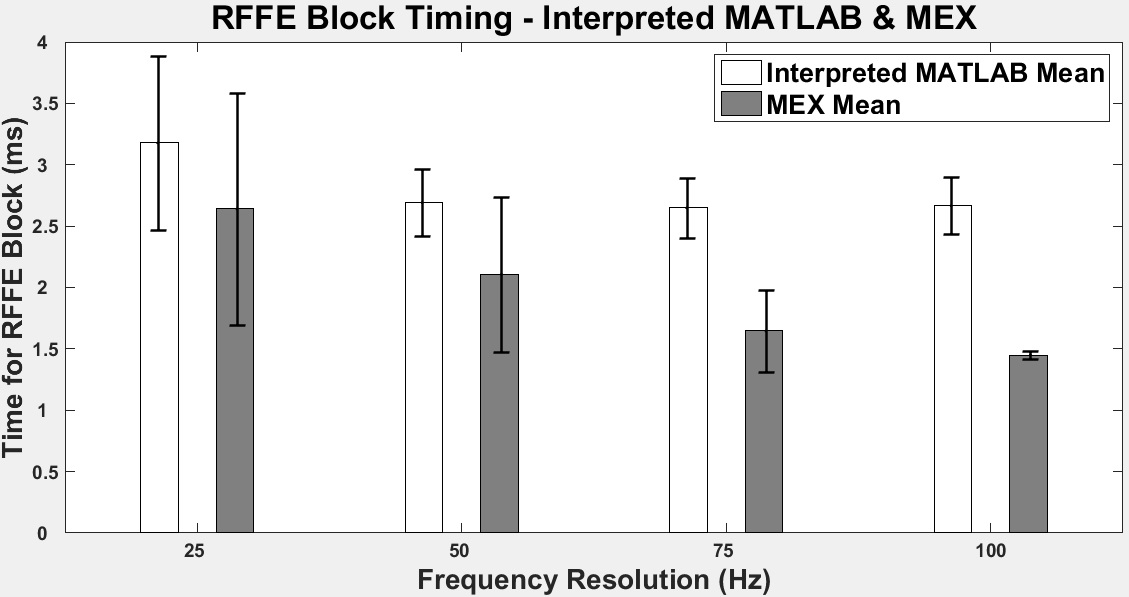}
\caption{RFFE block timing using interpreted MATLAB and MEX}
\label{fig_mexrffe}
\end{figure}

The timing of the RFFE block for various values of the frequency resolution parameter in interpreted MATLAB and C code compiled into MEX is shown in Fig. \ref{fig_mexrffe}. 
The addition of a FIR decimation step in the RFFE block reduces the sampling rate of the input for the subsequent coarse frequency offset estimation (CFOE). This reduction helps in increasing the frequency resolution, currently set at 100~Hz, which is the key parameter in controlling the execution time of CFOE. Further, we benefit from the improved accuracy of CFOE in that it corrects the signal so well that the later preamble detection block produces the correct synchronization delay to detect the start of DATA/ACK packet. The results clearly establish that average execution time for the RFFE block decreases with increase in frequency resolution. The reason for this is that CFOE uses progressively smaller FFT lengths. As before, the average execution time using MEX is generally smaller than using interpreted MATLAB. Also, the standard deviation for MEX results is always significantly less. Hence, MEX is a better option for the purpose of enforcing consistent RFFE execution times, which is required for slot-time synchronized operations. 
\subsection{Two Node Performance (1 DTx and 1 DRx)}
Link layer contention resolution and other MAC layer functions depends on the ability to reliably generate alternating DATA-ACK packets between the sender and receiver. In this regard, determining the performance of this basic link is important.

Packet error rate (PER) and bi-directional link latency are key performance indicators of the two node system. Of particular interest is the performance of the system when the transmit power level of the DTx is decreased below standard levels. The DTx was set up to send IEEE 802.11b compliant packets each with a large payload of random binary bits (2012 octets). The DRx receives the packet, checks for the correctness of the header information and acknowledges the receipt of the DATA packet by transmitting an ACK. The experiment was designed to be statistically significant, and hence, 100 packets were transmitted for each of the 5 different transmit gain settings. The results were averaged over 5 runs.

The experimental setup involved two host computers, both running MATLAB R2015b on a Ubuntu OS environment, each interfaced via the Ethernet cable to a USRP N210. The devices are configured to be DTx and DRx respectively and are kept about a meter apart.

\subsubsection{Packet Error Rate}
A packet is in error if the ACK for the same is not received in time by the DTx. This could mean that either the packet could not be decoded properly by the DRx or that the ACK was corrupted or lost while in transit to the DTx. An ideal system must recover quickly from such errors and, best trade-off PER and bi-directional link latency. PER is measured on average in percentage reflecting how many packets might be received in error for every 100 packets sent.
\begin{figure}[!t]
\centering
\includegraphics[width=3.5in]{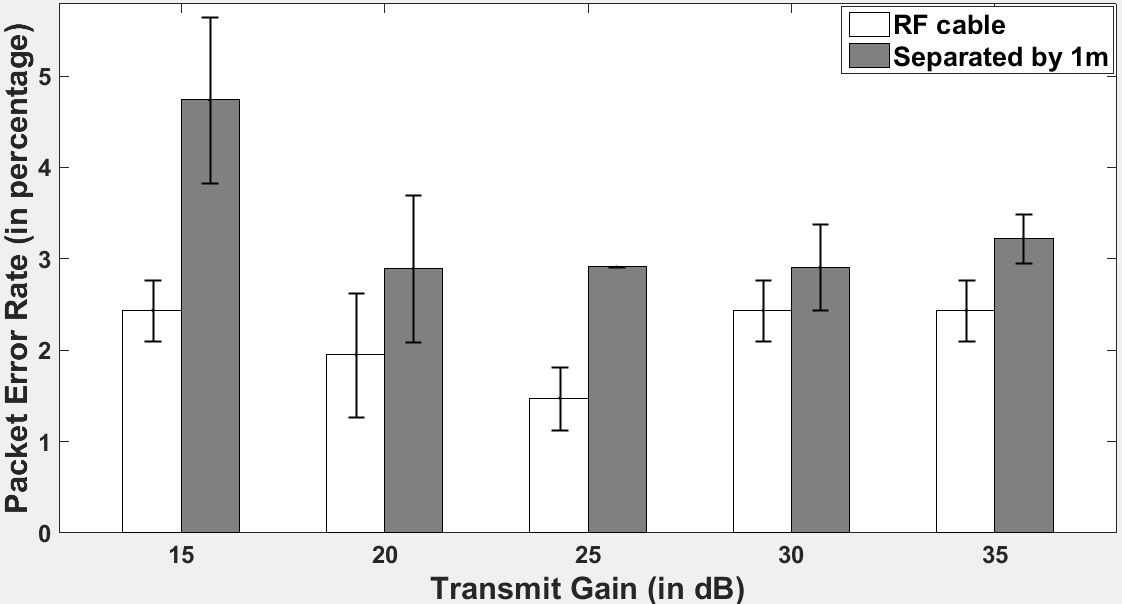}
\caption{Two Node Performance: Packet Error Rate}
\label{fig_PacketErrorRate}
\end{figure}
\subsubsection{Bi-directional Link Latency}
Bi-directional link latency is the average time taken by the DTx between sending a DATA packet and receiving the corresponding ACK packet. The bi-directional link latency includes any delay resulting from retransmissions accounting either for loss of DATA packet or ACK packet. Note that since the MAC layer code runs during the course of the experiment, the bi-directional link latency includes the DIFS duration and the random backoff period both set at 20~ms. The MAC layer functionality however is largely dormant in the 2 node case due to the lack of contention. Bi-directional link latency is averaged for a packet in seconds.
\begin{figure}[!t]
\centering
\includegraphics[width=3.5in]{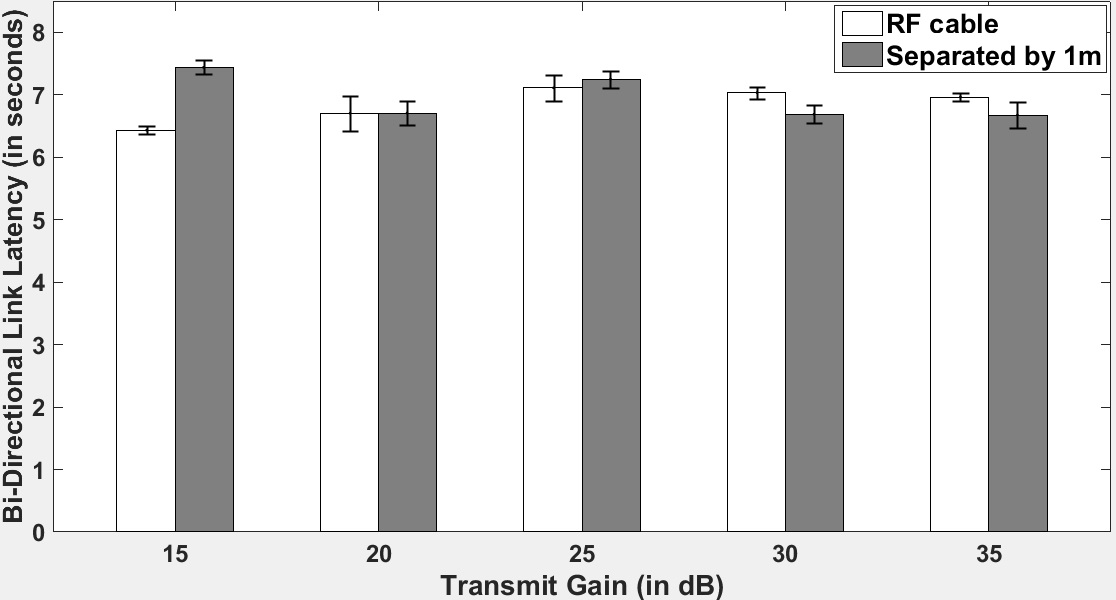}
\caption{Two Node Performance: Bi-directional Link Latency}
\label{fig_Latency}
\end{figure}

In the two node system, increasing DIFS and backoff time practically has no effect on the packet error rate due to lack of contention. However, increasing DIFS and backoff time also increases link latency by the same amounts. It should be noted that in the specifications, DIFS and contention window slot time are both fixed constants. 
\subsection{Profile of Time Elapsed in DTx States}
\begin{figure}[h]
\centering
\includegraphics[scale=0.25]{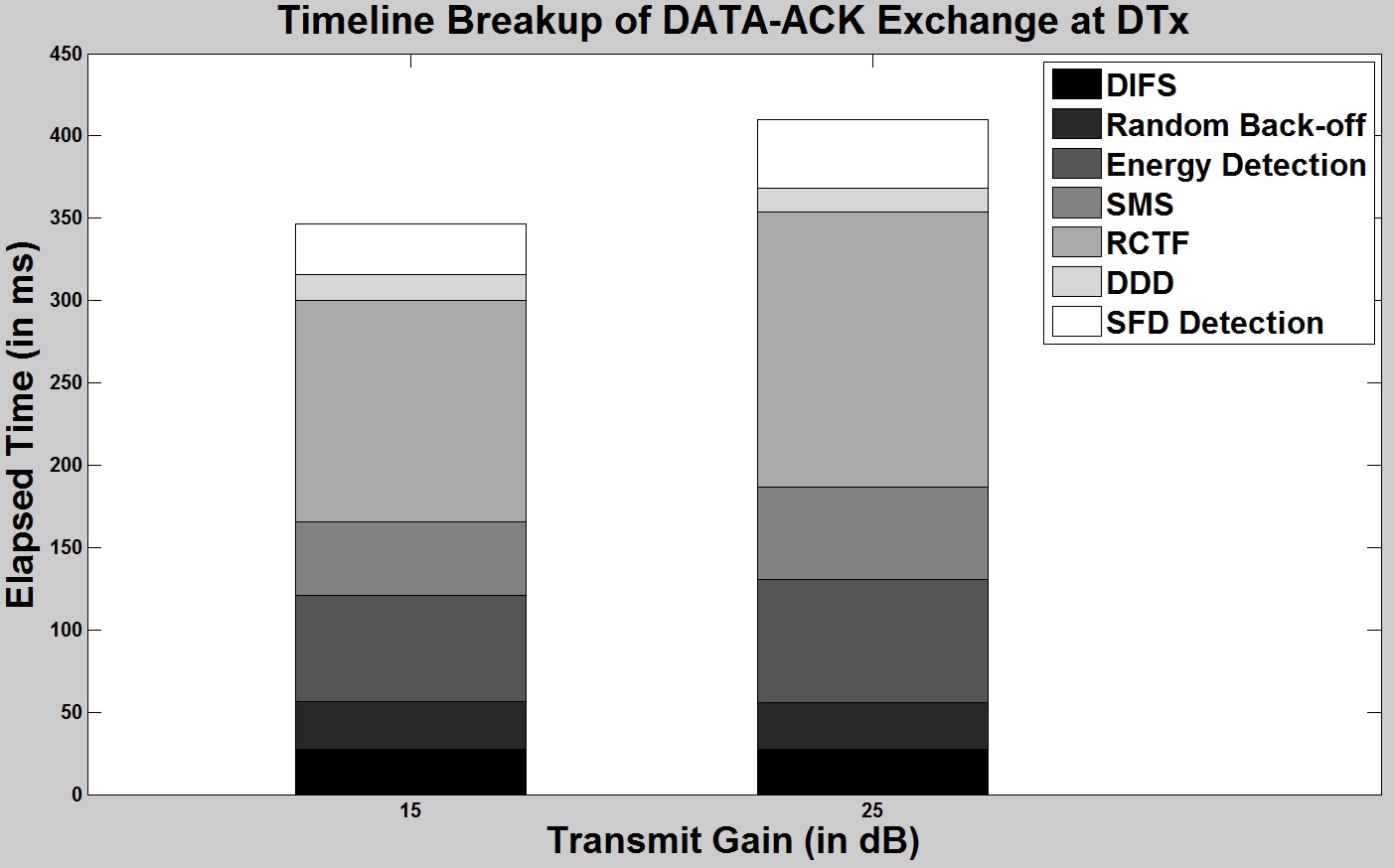}
\caption{Timeline Breakup of DATA-ACK Packet Exchange at DTx}
\label{fig_timelineBreakup1}
\end{figure}

\begin{figure}[h]
\centering
\includegraphics[scale=0.25]{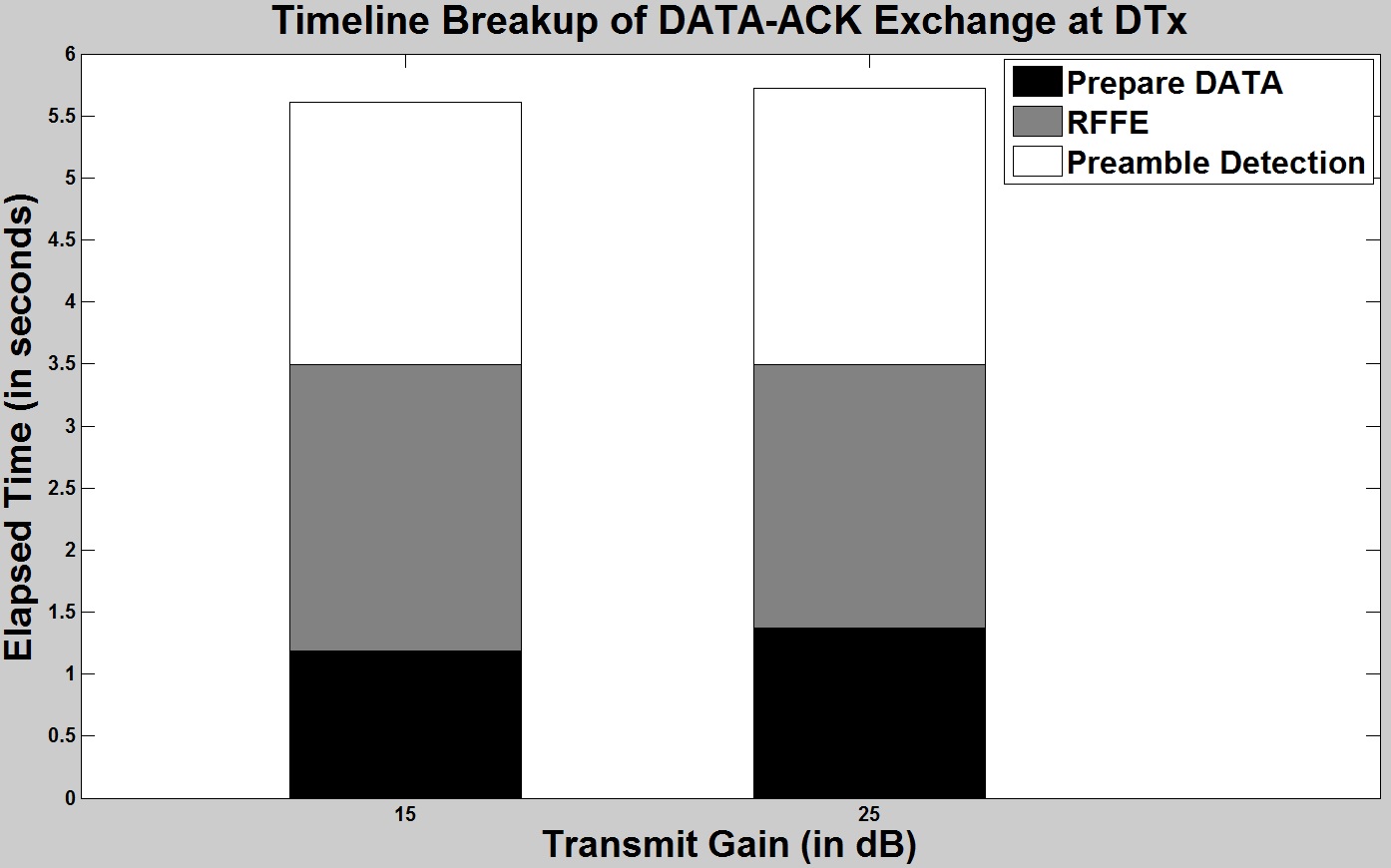}
\caption{Timeline Breakup of DATA-ACK Packet Exchange at DTx}
\label{fig_timelineBreakup2}
\end{figure}

At the DTx, we measured the time elapsed in each state for a DATA-ACK packet exchange. The stacked plots shown in Fig. \ref{fig_timelineBreakup1} and Fig. \ref{fig_timelineBreakup2} show the breakdown of the time spent in each substate. The plot at the top shows the small contributors to the overall processing time, and the one at the bottom shows the large contributors. Both the plots are part of the same DATA-ACK packet exchange and are separated for clarity. Note that (1) the time spent in the MAC portion of the code includes the time elapsed to detect energy in the channel continually together with the DIFS and random backoff duration, and (2) the time taken to send the IEEE 802.11b DATA packet includes the time to prepare the packet.

From Fig. \ref{fig_PacketErrorRate} and Fig. \ref{fig_Latency}, we can infer that the 2 node experiments show that the system guarantees a consistent $\leq$ 5\% packet error rate and approximately 7~seconds of bi-directional link latency (DATA-ACK packet exchange inclusive of the MAC functions) over a wide range of transmit gains (15-30~dB). Importantly, varying the distance between the 2 nodes does not significantly affect performance. Even moving the 2 nodes farther apart while still in line-of-sight (e.g. by 15 meters), the PER and bi-directional link latency stayed consistent. However, the presence of many metallic surfaces, such as in our lab setting, give rise to multi-path reflections that can be strong and result in packet errors. The fact that the performance was significantly better when the nodes were connected by RF cables confirms the case.

Keeping the packet sizes identical (DATA and ACK are 2072 octets and 40 octets long respectively), the standard off-the-shelf devices, operating at standard specified timings, the link latency $L_{std-link}$ (neglecting media contention, backoff times, and retransmissions) can be computed using Equation \ref{eqn_stdlinklat}. TxDATA and TxACK represent the elapsed time (in microseconds) to transmit a DATA packet and an ACK packet (at 1Mbps) respectively.
\begin{equation}
\label{eqn_stdlinklat}
\begin{split}
L_{std-link} & = DIFS + TxDATA + SIFS + TxACK \\
& = 50\mu s+(2072\times8) \mu s+10 \mu s+(40\times8) \mu s\\
& =16956 \mu s= 16.956 ms
\end{split}
\end{equation}
Comparing this to $t_{radio}$ in equation (\ref{eqn_tf}), we see that the link latency is in the same order as our slot time. Owing to hardware constraints, packet exchanges in standard devices are in the order of milliseconds while exchanges in this system are in the order of seconds. However, we argue that this is acceptable because our system adds the feature of software definition, which requires additional time for execution. 

\subsection{Three Node Experimental Setup (2 DTxs and 1 DRx)}
Given that without the MAC layer, the DATA/ACK packet collisions and the link latencies will be unacceptably high, we performed experiments to assess the MAC performance with a set of 3 USRPs (three nodes: 2 DTxs and 1 DRx). To that end, we implemented MAC functions to distinguish the two links and fine-tuned the MAC/PHY parameters of the system. We expect to see increased bi-directional link latency and PER as the DTxs contend to gain access to the channel leading to packets collisions and subsequent retransmits.

\begin{figure}[!t]
\centering
\includegraphics[width=3in]{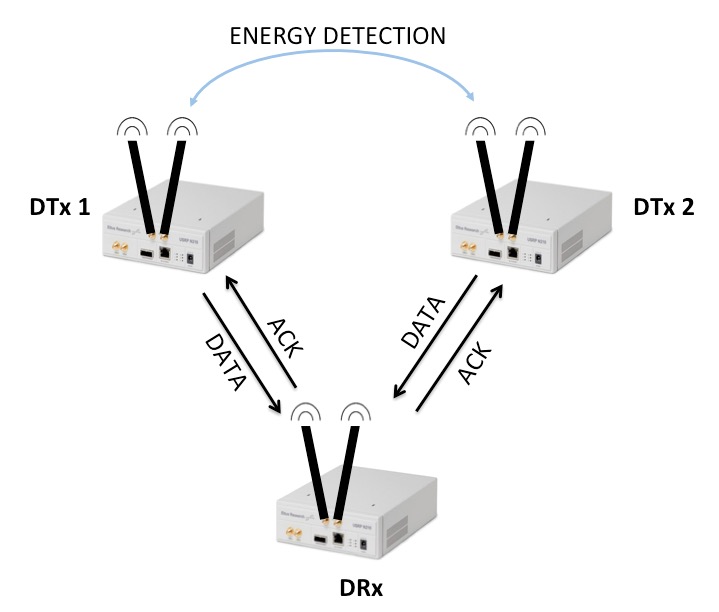}
\caption{Three Node System with 2 DTxs and 1 DRx}
\label{fig_ThreeNodeSystem}
\end{figure}

In our 2 node experiments, we confirmed that for a wide range of transmit gains, the performance remains consistent. We now have two independent links incident on one shared DRx, and hence, we do not expect to see much difference in the performance of the two links when varying the transmit gains here in the 3 node case. Instead, we measured bi-directional Link Latency and Packet Error Rate for DATA-ACK packet exchange in the two links as shown in Fig. \ref{fig_ThreeNodeSystem} by varying the payload size in the DATA packet. Essentially, the experiments let us compare the individual performances of the two links and further establish the MAC layer's role in enforcing fairness among the DTxs in accessing the channel.
\subsubsection{Implemented MAC functions}
\begin{figure}[!t]
\centering
\includegraphics[width=3.3in]{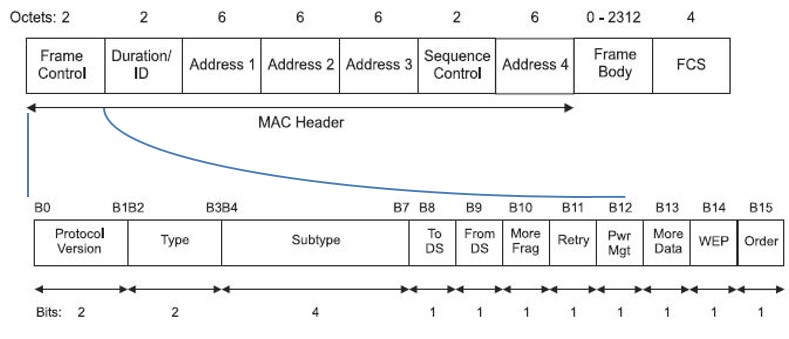}
\caption{MAC Header - DATA packet \cite{ieee80211}}
\label{fig_macDataHeader}
\end{figure}

\begin{figure}[!t]
\centering
\includegraphics[width=1.8in]{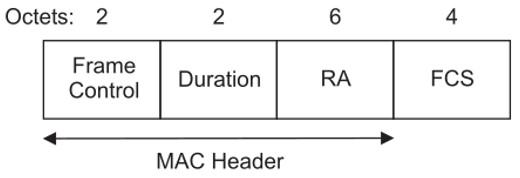}
\caption{MAC Header - ACK packet \cite{ieee80211}}
\label{fig_macAckHeader}
\end{figure}

The MAC header format for DATA and ACK shown in Fig. \ref{fig_macDataHeader} and Fig. \ref{fig_macAckHeader} respectively will aid in discussion of the MAC layer functions \cite{ieee80211}.

The DRx determines the DTx address from the MAC header of the received DATA packet and sends out an ACK addressed to that DTx.  Furthermore, the DRx can reject DATA packets not addressed to it. Note that steps right from preamble detection, SFD detection, all the way up to reading into the IP address of the DTx from the MAC header, are carried out at the DRx, preceding the rejection of that DATA packet. On the other hand, the DTxs can determine the DRx from the MAC header of the received ACK and can go on to either accept or reject the ACK based on the IP Address. Previously, we had the DTx re-transmitting DATA packet only towards lost ACKs. Clearly, these are the MAC functions necessary for scaling up the system, enabled by reading into the MAC header of the DATA/ACK packet.
\subsubsection{MAC parameters}
We learned from our initial set of experiments that the DATA/ACK packet processing in the host machine takes significantly more time compared to time taken in transmitting a DATA packet. This is expected as most SDRs use a host computer for processing. Also, the SIFS duration, set in the order of microseconds in commercial products, imposes a time constraint in most SDRs that is difficult to achieve. The reason is that the latency for the signal to move back and forth from the radio to the host exceeds the SIFS duration requirements. The standard specifies the constants as follows: Slot-time = 20 $\mu$s, SIFS = 10 $\mu$s, DIFS = SIFS + 2 x Slot-time = 50 $\mu$s.

The experiments helped us fine-tune the DIFS duration (which the standard specifies be greater than SIFS), random backoff duration, and ACK timeout duration towards fewer packet collisions. As a result, we performed our experiments with DIFS duration, minimum contention window, and ACK timeout duration set at 0.75, 0.5, and 5.0~seconds, respectively.

\subsubsection{Picking the Energy Threshold}
Three node performance relies heavily on the energy detection step at both the DTxs. Accuracy of energy detection is critical and it requires the energy threshold be carefully picked at both the DTxs, enabling each DTx to back off as soon as they sense another DTx transmit, and subsequently transmit at the right instants of time, thereby keeping the packet errors and bi-directional link latency to a desired minimum. Additionally, it enforces fairness towards channel access among the DTxs. 

The receive gain set at the DTx and the inter-node distances (1 meter in our experiments) affect the magnitude of the energy threshold. A value close to and slightly above the \emph{noise floor} set as the energy threshold will not work as intended, as a power-cycle of the USRP changes it. Also, an energy threshold set at a large value might not allow the DTxs to sense each other transmitting due to rapidly fluctuating RF power output despite the AGC. Therefore, each DTx may not backoff at the right instants, leading to collisions at the DRx. However, by picking a small enough energy threshold, which is enough to detect signal energy over channel noise, we could make each DTx sensitive enough to sense the other DTx transmitting and backoff fairly well, thereby reducing packet retransmissions. 

\subsection{Three Node Performance: Experimental Results}
\begin{figure}[!t]
\centering
\includegraphics[width=3.5in]{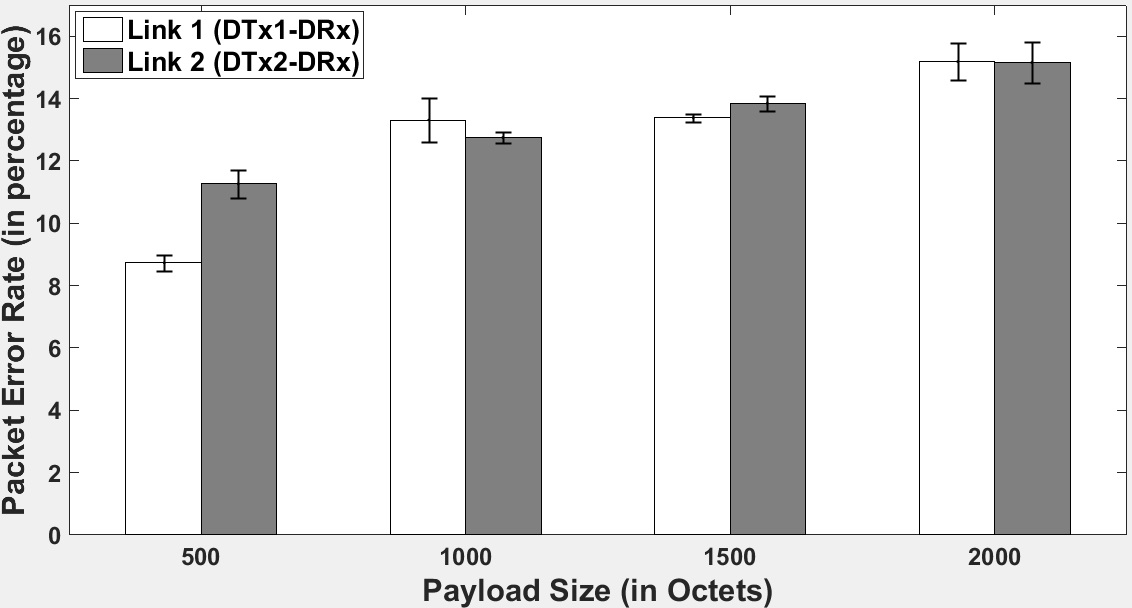}
\caption{Three Node Performance - Packet Error Rate of the Links}
\label{fig_3NodePayloadPktErr}
\end{figure}

\begin{figure}[!t]
\centering
\includegraphics[width=3.5in]{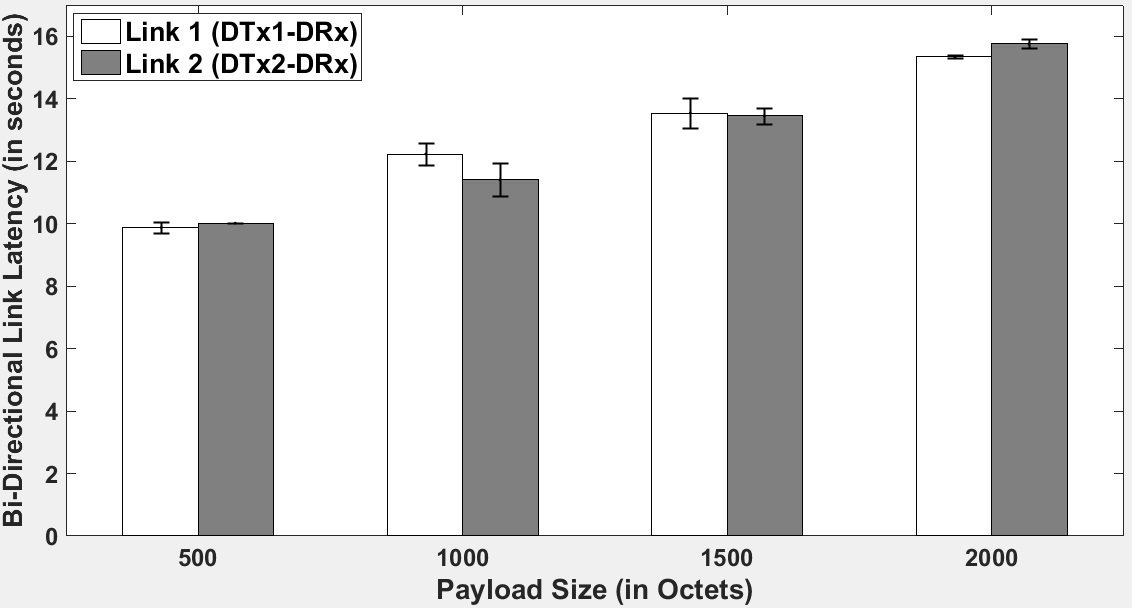}
\caption{Three Node Performance - Bi-directional Link Latencies}
\label{fig_3NodePayloadLatency}
\end{figure}

Packet error rate and bi-directional link latency for DATA-ACK packet exchanges in the two links varying the payload size in the DATA packet are shown in Fig. \ref{fig_3NodePayloadPktErr} and Fig. \ref{fig_3NodePayloadLatency}, respectively. Four different payload sizes, 500, 1000, 1500, and 2000 octets, were used for the experiment to measure 3 node performance.

Smaller payload sizes correspond to smaller packets and decreased time that the DTx is occupying the channel whereas larger payload sizes increases the likelihood of packet collisions. The link latency and the packet error rate in the latter is bound to increase as larger packets incur higher processing delay at the DRx and more collisions necessitating increased packet retransmits.

\subsubsection{Goodput}
Goodput, a performance measure used in computer networks, is the rate at which useful information bits traverse a link.
Goodput can be measured using equation (\ref{Goodput}),
\begin{equation} \label{Goodput}
\small Goodput = \frac{Total\ payload\ bits\ correctly\ decoded}{Average\  Bi\mbox{-}directional\ Link\ Latency} \\
\end{equation}
The average Goodput of the two bi-directional links computed using (\ref{Goodput}) are shown in Table \ref{table_goodput}. Notice that the goodput increases with the payload size. The reason for this is that the combined PHY and MAC header occupies a decreased fraction of the entire DATA packet as the payload size increases.

\begin{table}[h]
\renewcommand{\arraystretch}{1}
\caption{Average Goodput for Varying Payload Sizes}
\label{table_goodput}
\centering
\begin{tabular}{|c||c|c|}
\hline \bfseries Payload Size  & \bfseries Link 1 Goodput & \bfseries Link 2 Goodput \\
 (\#Octets) & (Kbps) & (Kbps) \\
%\hline 500  & 0.4050 & 0.3993\\
%\hline 1004 & 0.6568 & 0.7043\\
%\hline 1500 & 0.8862 & 0.8922\\
%\hline 2004 & 1.0450 & 1.0172\\
\hline 500  & 0.41 & 0.40\\
\hline 1004 & 0.66 & 0.70\\
\hline 1500 & 0.89 & 0.89\\
\hline 2004 & 1.05 & 1.02\\
\hline
\end{tabular}
\end{table}

In the three node system, when there is a symmetric increase in DIFS and backoff time at the two DTxs, then the system will remain fair with reduced contention, resulting in fewer packet errors. However, the goodput decreases as link latency increases. Also note that the standard specifies the DIFS and the contention window slot time be fixed constants.

\begin{figure}[!t]
\centering
\includegraphics[width=3in]{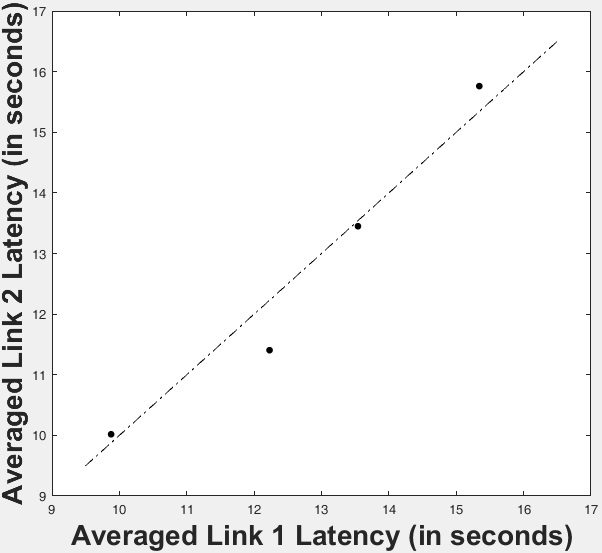}
\caption{MAC Layer Fairness - Averaged Link Latencies}
\label{fig_3NodeFairnessAvgLatency}
\end{figure}
\subsubsection{Fairness}
The line shown in Fig. \ref{fig_3NodeFairnessAvgLatency} is representative of an ideal system, in  which the two DTxs access the channel equally often, such that their bi-directional link latencies are identical. Fairness is an important feature for the system to possess, and is brought about by the MAC protocol.

Notice that the latencies of the two links deviate by only a small amount from the ideal line for varying payload sizes. This result establishes the role and efficacy of the MAC layer in enabling and enforcing fairness among the two DTxs when accessing the common channel. 
\section{Conclusions}
\label{sec_conclusion}
Building our system around the concept of state-action based design and slot-time synchronized operations helped combine and realize the PHY and MAC layer that is IEEE 802.11b standard compliant. In addition, the system allows the user reconfigure the parameter values as needed.
Using the MATLAB Coder to automatically generate MEX functions is beneficial in improving the speed consistency of our system blocks, most notably RFFE, which can vary its frequency resolution parameter.
This work provides a testbed to experiment with new MAC protocols beyond that specified in the IEEE 802.11b standard. The state machine design enables modularity of code base and should allow for extensibility by the community. The three node system remains fair to the two bi-directional links for varying payload sizes in the DATA packet. Through our experiments we have established the role and efficacy of the implemented MAC layer towards mitigating packet collisions and enforcing fairness among DTxs in accessing a common channel.

There were a number of difficulties during the implementation that we had to overcome. Foremost, we had trouble realizing slot-synchronized operations, one of the most crucial issues in real-time testbeds. Second, it was difficult to pick the right energy threshold to deal with a variable noise floor due to environmental noise effects. Finally, our system required a thorough calibration step prior to running experiments. The minimum receive gain settings at the devices are always different. While performing the experiments, we took care to isolate the experimental setup from highly reflective metallic surfaces and external transmissions, as is typical in a lab environment.

These experimental results have provided us with performance benchmarks that will focus future work on further optimization and sophistication of the MATLAB-based MAC layer. Also, as part of our future work, we plan use this framework to perform evaluation studies on the co-existence of LTE and 802.11 Wi-Fi networks.

\appendices
\section*{Acknowledgments}
This work is supported by MathWorks under the Development-Collaboration Research Grant A\#: 1-945815398. We would like to thank Mike McLernon and Ethem Sozer for their continued support on this project. We would also like to thank Taylor Skilling for his support with the experiments. 

\ifCLASSOPTIONcaptionsoff
  \newpage
\fi
\bibliographystyle{IEEEtran}    
\bibliography{IEEEabrv,mybibfile}
\vspace{-12 mm}
\begin{IEEEbiography}[{\includegraphics[width=1in,height=1.5in,clip,keepaspectratio]{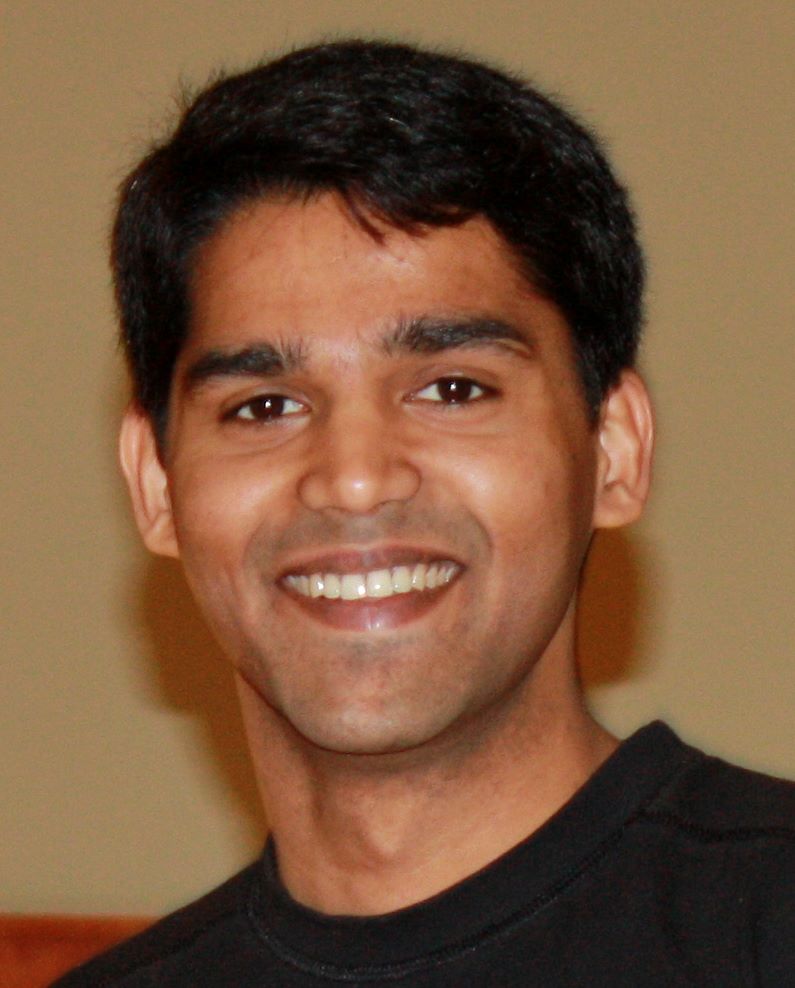}}]{Ramanathan Subramanian}
Ram is a Ph.D. candidate in the Department of Electrical and Computer Engineering at Northeastern University, Boston advised by Prof. Kaushik Chowdhury. He completed his Masters in Computer Science and Automation at Indian Institute of Science, Bangalore, India. His current efforts are focused on implementing the MAC layer functionality for MATLAB-based SDR on the USRP hardware. He then will progress to research MAC mechanisms that allow LTE to co-exist with Wi-Fi in the unlicensed spectrum. 
\end{IEEEbiography}
\vspace{-12 mm}
\begin{IEEEbiography}[{\includegraphics[width=1in,height=1.5in,clip,keepaspectratio]{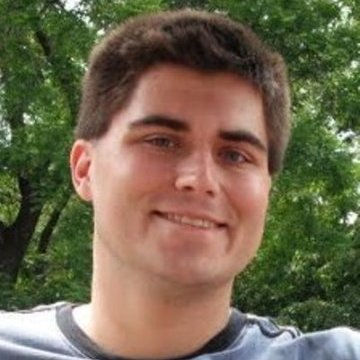}}]{Benjamin Drozdenko}
Ben is a Ph.D. candidate at Northeastern University, the university-wide MathWorks TA, and a graduate research assistant in implementation of a MATLAB-based Cognitive Radio framework, co-advised by Prof. Chowdhury and Prof. Leeser. From 2008 to 2014, Ben worked for MathWorks, Inc., producers of MATLAB and Simulink, where he wrote technical documentation and examples for several Signal Processing and Communications area products and traveled to customers' sites to deliver training courses. From 2004 to 2008, Ben worked as a Systems Engineer at Raytheon Integrated Defense Systems, focusing on ground-based radar for ballistic missile defense.
\end{IEEEbiography}
\newpage
\vspace{-12 mm}
\begin{IEEEbiography}[{\includegraphics[width=1in,height=1.5in,clip,keepaspectratio]{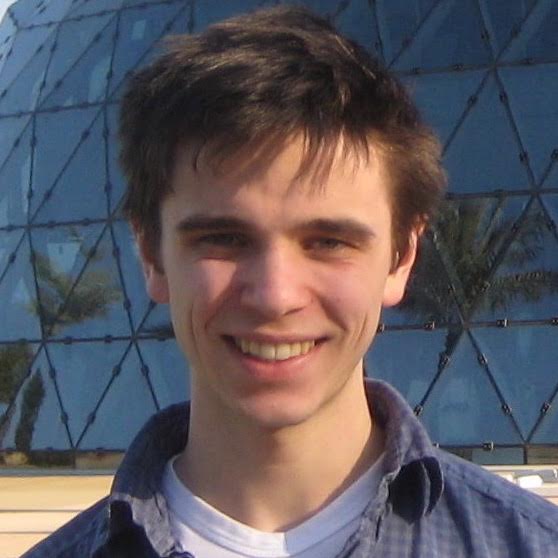}}]{Eric Doyle}
Eric is an undergraduate student at Northeastern University studying Electrical and Computer Engineering and has been a research assistant in Prof. Kaushik Chowdhury's GENESYS Laboratory since August 2015. His research interests are in communication systems and acoustics. He has been vice president of the Acoustical Society of America chapter at Northeastern since 2014 and has held coop positions at Bose Corporation and Insulet Corporation. In his free time he enjoys practicing violin. 
\end{IEEEbiography}
\vspace{-10 mm}
\begin{IEEEbiography}[{\includegraphics[width=1in,height=1.5in,clip,keepaspectratio]{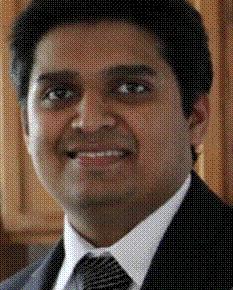}}]{Rameez Ahmed}
Rameez is a Ph.D. candidate in the Department of Electrical and Computer Engineering at Northeastern University, Boston. He works under the guidance of Prof. Milica Stojanovic in the field of underwater acoustic communication. His research interests are in digital communication, wireless communication and network protocols. In his free time he enjoys cooking and traveling and has an ardent love for cars and electronic gadgets.
\end{IEEEbiography}
\vspace{-10 mm}
\begin{IEEEbiography}[{\includegraphics[width=1in,height=1.5in,clip,keepaspectratio]{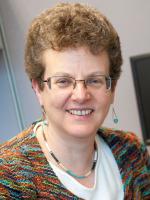}}]{Miriam Leeser}
Miriam Leeser is a Professor in Electrical and Computer Engineering at Northeastern where she is head of the Reconfigurable and GPU Computing Laboratory.  She received her BS degree in electrical engineering from Cornell University, and Diploma and PhD degrees in computer science from Cambridge University in England. She was on the faculty at Cornell University's Department of Electrical Engineering before coming to Northeastern, where she is a member of the Computer Engineering research group. Her research includes using heterogeneous architectures for signal and image processing applications including wireless communications as well as implementing computer arithmetic and verifying critical applications. She is a senior member of the IEEE and of the ACM.
\end{IEEEbiography}
\vspace{-10 mm}
\begin{IEEEbiography}[{\includegraphics[width=1in,height=1.5in,clip,keepaspectratio]{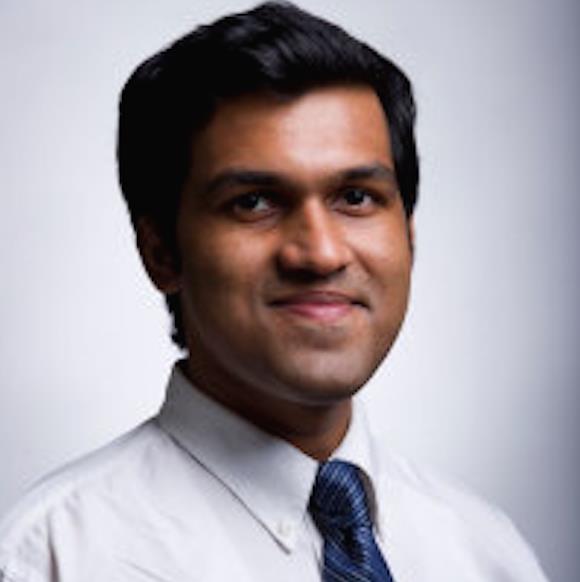}}]{Kaushik Chowdhury}
Kaushik Roy Chowdhury is Associate Professor in the Electrical and Computer Engineering Department at Northeastern University. He was earlier Assistant Professor in the same university from 2009-2015. He received his Ph.D. from the Georgia Institute of Technology in August 2009, under Prof. Ian F. Akyildiz and M.S. from the University of Cincinnati in 2006, advised by Prof. Dharma Agrawal. Prof. Chowdhury is the winner of the NSF CAREER award in 2015. He received the best paper award at the IEEE International Conference on Communications (ICC) in 2013, 2012 and 2009, as well as the best paper award at the International Conference on Computing, Networking and Communications (ICNC) in 2013. He is the current Chair of the IEEE Technical Committee on Simulation and is Sr. Member of the IEEE. He serves as the area editor for the following journals: Elsevier Ad Hoc Networks, Elsevier Computer Communications, and EAI Transactions on Wireless Spectrum. 
\end{IEEEbiography}
\vfill
\end{document}